\documentclass[final,3p,times,twocolumn,authoryear]{elsarticle}

\pdfoutput=1

\usepackage[noend]{distribalgo}
\usepackage{algorithm}
\usepackage{times}
\usepackage{amsmath}
\usepackage{color}
\usepackage{amssymb}
\usepackage{url}
\usepackage{comment}

\newcommand{\mv}[1]{\ensuremath{\operatorname{\mathit{#1}}}}
\definecolor{dark}{gray}{.6}
\newcommand{\bc}[1]{\textcolor{dark}{#1}}

\newtheorem{props}{Proposition}

\usepackage{amssymb}
\journal{Journal of Parallel and Distributed Computing}

\begin{document}

\begin{frontmatter}



\title{Ring Paxos: High-Throughput Atomic Broadcast \tnoteref{tlabel}}
\tnotetext[tlabel]{This paper is an extended version of ``Ring Paxos: A High-Throughput Atomic Broadcast Protocol", published at DSN 2010 by the same authors. This paper extends our previous work in four directions: (a)~it proposes a new protocol well suited for environments in which ip-multicast is not available, including its implementation and detailed performance results; (b)~it extends the design considerations for the algorithms we propose; (c)~it adds flow control to Ring Paxos and evaluates its effectiveness; (d)~it evaluates both protocols when data is persisted through synchronous disk writes.}

%\title{Title\tnoteref{This paper is an extended version of ``Ring Paxos: A High-Throughput Atomic Broadcast Protocol", published at DSN 2010 by the same authors. This paper extends our previous work in four directions: (a)~it proposes a new protocol well suited for environments in which ip-multicast is not available, including its implementation and detailed performance results; (b)~it extends the design considerations for the algorithms we propose; (c)~it adds flow control to Ring Paxos and evaluates its effectiveness; (d)~it evaluates both protocols when data is persisted through synchronous disk writes.}}

%\address{}

\author{Parisa Jalili Marandi\fnref{usi}}
\author{Marco Primi\fnref{apple}}
\author{Nicolas Schiper\fnref{cornell}}
\author{Fernando Pedone\fnref{usi}}
\fntext[usi]{Faculty of Informatics, University of Lugano, Switzerland.}
\fntext[apple]{Apple Inc, San Francisco, USA.}
\fntext[cornell]{Computer Science Department, Cornell University, USA.}
%\ead{nn@,,,}

\begin{abstract}
Atomic broadcast is an important communication primitive often used to implement state-machine replication. 
Despite the large number of atomic broadcast algorithms proposed in the literature, few papers have discussed how to turn these algorithms into efficient executable protocols. 
This paper focuses on a class of atomic broadcast algorithms based on Paxos, with its corresponding desirable properties: safety under asynchrony assumptions, liveness under weak synchrony assumptions, and resiliency-optimality.
The paper presents two protocols, M-Ring Paxos and U-Ring Paxos, derived from Paxos. 
The protocols inherit the properties of Paxos and can be implemented very efficiently. 
We report a detailed performance analysis of M-Ring Paxos and U-Ring Paxos and compare them to other atomic broadcast protocols.

\end{abstract}


\end{frontmatter}


%\section{}
%\label{}

\section{Introduction}
\label{sec:intro}

State-machine replication is a fundamental approach to building fault-tolerant distributed systems~\cite{Lam78,Sch93}. 
The idea is to replicate a service so that faulty replicas do not prevent operational replicas from executing service requests. 
State-machine replication can be decomposed into two requirements regarding the dissemination of requests to replicas: 
(i)~every nonfaulty replica receives every request and (ii)~no two replicas disagree on the order of received requests. 
These two requirements are often encapsulated in a group communication primitive known as atomic broadcast or total-order broadcast~\cite{HT93}.

Being at the core of state-machine replication, atomic broadcast has an important impact on the overall performance of a replicated service and a lot of effort has been put into designing efficient atomic broadcast algorithms~\cite{DUS04}. 
Fewer papers, however, have discussed how to turn these algorithms into efficient systems. 
We focus on a class of atomic broadcast algorithms based on Paxos~\cite{Lam98}.
Paxos has important properties: it is safe under asynchrony assumptions, live under weak synchrony assumptions, and resiliency-optimal, that is, it requires a majority of non-faulty processes to ensure progress.
We are interested in Paxos implementations that aim at high throughput.
%Although this paper explicitly addresses atomic broadcast in a clustered environment, some of our design considerations are general enough to be useful in the development of  other distributed protocols.

We are interested in efficiency as a measure of throughput. 
More precisely, we define the efficiency of an atomic broadcast protocol as the rate between its maximum achieved throughput per receiver, and the nominal transmission capacity of the system per receiver. 
For example, a protocol that allows a receiver to deliver up to 500 Mbps of application data in a system equipped with a gigabit network has efficiency 0.5 or 50\%. 
An efficient protocol would have high efficiency (e.g, 90\% or more), ideally independent of the number of receivers.
%An ideal protocol would have efficiency near 1; moreover, this efficiency would not be dependent on the number of receivers. 
Due to inherent limitations of an algorithm, implementation details, and various overheads (e.g., added by the network layers), typical atomic broadcast protocols are not ideal according to this metric.

This paper presents M-Ring Paxos and U-Ring Paxos, two efficient atomic broadcast protocols: they have efficiency above 90\%, which in some cases does not depend on the number of receivers. 
The protocols are based on Paxos and inherit many of its characteristics: they are safe under asynchrony assumptions, live under weak synchrony assumptions, and resiliency-optimal. 
We revisit Paxos in light of a number of optimizations and from these we derive M-Ring Paxos and U-Ring Paxos.
Although our optimizations mostly address atomic broadcast in a clustered environment, some of our design considerations are general enough to be useful in the development of other distributed protocols.

M-Ring Paxos uses ip-multicast and unicast (i.e., UDP). Network-level multicast is a powerful communication primitive to propagate messages to a set of processes in a cluster since it delegates to the interconnect (i.e., ethernet switch) most of the communication. 
However, ip-multicast is subject to message losses, mostly due to buffer overflow---when the receiver is not able to consume messages at the rate they are transmitted. 
M-Ring Paxos uses a single ip-multicast stream to disseminate messages and thus benefit from the throughput that ip-multicast can provide without falling prey to its shortcomings. To evenly balance the incoming and outgoing communication needed to totally order messages, M-Ring Paxos places $f+1$ nodes in a logical ring, where $f$ is the number of tolerated faulty processes.

U-Ring Paxos uses unicast communication only (i.e., TCP) and disposes all processes in a ring. U-Ring Paxos is not the first atomic broadcast protocol to place nodes in a logical ring (e.g., Totem~\cite{amir1995totem}, LCR~\cite{Guerraoui2010}, and the protocol in~\cite{ESU04} have done it before), but it is the first to achieve very high throughput, almost constant with the number of receivers, with the resilience and synchrony assumptions of Paxos.

We implemented M-Ring Paxos and U-Ring Paxos and compared their performance to other Paxos protocols. 
In particular, our Paxos-derived protocols can reach an efficiency of more than 90\% in a gigabit network. 
M-Ring Paxos has low delivery latency, below 5 msec, which remains approximately constant with an increasing number of receivers (up to 25 receivers in our experiments).
Previous implementations of the Paxos protocol, based either on ip-multicast only or on unicast only, have efficiency below 90\%. 
The only other atomic broadcast protocol that achieves high efficiency we know of is LCR~\cite{Guerraoui2010}, a pure ring-based protocol. 
But it relies on stronger synchrony assumptions than Paxos. 
U-Ring Paxos and LCR have latency that depends on the number of processes in the ring.

Briefly, this paper makes three contributions: 
First, it proposes novel atomic broadcast algorithms for clustered networks derived from Paxos. 
Second, it describes implementations of these algorithms. 
Third, it analyses their performance and compares them to other atomic broadcast protocols.
The remainder of the paper is structured as follows.
Section~\ref{sec:moddef} describes our system model.
Section~\ref{sec:paxos} reviews Paxos.
Section~\ref{sec:rpaxos} presents the observations our protocols are based on and introduces
M-Ring Paxos and U-Ring Paxos.
Section~\ref{sec:rwork} comments on related work.
Section~\ref{sec:perf} evaluates the performance of the protocols and
compares them quantitatively to a number of other protocols.
Section~\ref{sec:final} concludes the paper.
The Appendix contains a correctness argument for our protocols.

\section{Model and definitions}
\label{sec:moddef}

\subsection{System model}
\label{sec:model}

We assume a distributed system model in which processes communicate by exchanging messages. Processes can fail by crashing but never perform incorrect actions (i.e., no Byzantine failures).
Communication can be based on \emph{unicast}, through the primitives \emph{send}$(p,m)$ and \emph{receive}$(m)$, and \emph{multicast}, through the primitives \emph{ip-multicast}$(g,m)$ and \emph{ip-deliver}$(m)$, where $m$ is a message, $p$ is a process, and $g$ is the group of processes $m$ is addressed to. Messages can be lost but not corrupted. In the text we refer sometimes to multicast messages as \emph{packets}.

Our protocols, like Paxos, ensure safety under both asynchronous and synchronous execution periods. The FLP impossibility result~\cite{FLP85} states that in an asynchronous system consensus and atomic broadcast cannot be both safe and live. We thus assume the system is \emph{partially synchronous}~\cite{DLS88}, that is, it is initially asynchronous and eventually becomes synchronous. The time when the system becomes synchronous is called the \emph{Global Stabilization Time (GST)}~\cite{DLS88}, and it is unknown to the processes.

Before GST, there are no bounds on the time it takes for messages to be transmitted and actions to be executed. After GST, such bounds exist but are unknown. Moreover, in order to prove liveness, we assume that \emph{after} GST all remaining processes are \emph{correct}---a process that is not correct is \emph{faulty}. A correct process is operational ``forever" and can reliably exchange messages with other correct processes. Notice that in practice, ``forever" means long enough for consensus to terminate.

\subsection{Consensus and atomic broadcast}

Consensus and atomic broadcast are two distributed agreement problems at the core of state-machine replication. The problems are related: atomic broadcast can be implemented using a sequence of consensus executions~\cite{CT96}. Consensus is defined by the primitives \emph{propose}$(v)$ and \emph{decide}$(v)$, where $v$ is an arbitrary value; atomic broadcast is defined by the primitives \emph{broadcast}$(m)$ and \emph{deliver}$(m)$, where $m$ is a message.

Consensus guarantees that (i)~if a process decides $v$ then some process proposed $v$; (ii)~no two processes decide different values; and (iii)~if one or more correct processes propose a value then eventually some value is decided by all correct processes.
Atomic broadcast guarantees that (i)~if a process delivers $m$, then all correct processes deliver $m$; (ii)~no two processes deliver  messages in different orders; and (iii)~if a correct process broadcasts $m$, then all correct processes deliver $m$.

\section{Paxos}
\label{sec:paxos}

Paxos is a fault-tolerant consensus algorithm intended for state-machine replication~\cite{Lam98}.
We describe next how a value is decided in a single instance of consensus.

Paxos distinguishes three roles: \emph{proposers}, \emph{acceptors}, and \emph{learners}. A process can execute any of these roles, and multiple roles simultaneously. Proposers propose a value, acceptors choose a value, and learners learn the decided value. Hereafter, $N_a$ denotes the set of acceptors, $N_l$ the set of learners, and $Q_a$ a \emph{majority quorum} of acceptors (\emph{m-quorum}), that is, a subset of $N_a$ of size $\lceil (|N_a|+1)/2 \rceil$.

The execution of one consensus instance proceeds in a sequence of \emph{rounds}, uniquely identified by a round number, a positive integer. For each round, one process, typically among the proposers or acceptors, plays the role of \emph{coordinator} of the round. To propose a value, proposers send the value to the coordinator.
The coordinator maintains two variables: (a)~\mv{c-rnd} is the highest-numbered round that the coordinator has started; and (b)~\mv{c-val} is the value that the coordinator has picked for round \mv{c-rnd}. The first is initialized to 0 and the second to null.

Acceptors maintain three variables: (a)~\mv{rnd} is the highest-numbered round in which the acceptor has participated, initially 0; (b)~\mv{v-rnd} is the highest-numbered round in which the acceptor has cast a vote, initially 0---it follows that $\mv{v-rnd} \leq \mv{rnd}$ always holds; and (c)~\mv{v-val} is the value voted by the acceptor in round \mv{v-rnd}, initially null.

Algorithm~1 provides an overview of Paxos. The algorithm has two phases. To execute Phase~1, the coordinator picks a unique round number \mv{c-rnd} greater than any value it has picked so far, and sends it to the acceptors (Task~1). Upon receiving such a message (Task~2), an acceptor checks whether the round proposed by the coordinator is greater than any round it has received so far; if so, the acceptor promises not to accept any future message with a round smaller than \mv{c-rnd}. The acceptor then replies to the coordinator with the highest-numbered round in which it has cast a vote, if any, and the value it voted for in this round. Notice that the coordinator does not send any proposal in Phase~1.
\begin{algorithm}

%\line(1,0){240}
\begin{distribalgo}[1]
\small

\STATE \textbf{Algorithm 1: Paxos} (for process $p$)
\vspace{1mm}
\STATE \emph{\underline{Task~1 (coordinator)}}
\INDENT{\textbf{upon} receiving value $v$ from proposer $P(v)$}
%\STATE{\textbf{upon} receiving value $v$ from proposer $P(v)$}
	\STATE increase \mv{c-rnd} to an arbitrary unique value
	\STATE \textbf{for all} $q \in N_a$ \textbf{do} send ($q$, (\textsc{Phase 1a}, \mv{c-rnd}))
\ENDINDENT

\vspace{1.5mm}
\STATE \emph{\underline{Task~2 (acceptor)}}
\INDENT{\textbf{upon} receiving (\textsc{Phase 1a}, \mv{c-rnd}) from coordinator}
	\IF{$\mv{c-rnd} > \mv{rnd}$}
		\STATE let \mv{rnd} be \mv{c-rnd}
		\STATE send (coordinator, (\textsc{Phase 1b}, \mv{rnd}, \mv{v-rnd}, \mv{v-val}))
	\ENDIF
\ENDINDENT
%\INDENT{\textbf{upon} receiving (\textsc{Phase 1b}, \mv{v-rnd}, \mv{v-val}) from $\lceil N_a/2 \rceil$\\
 %\hfill acceptors}
\vspace{1.5mm}
\STATE \emph{\underline{Task~3 (coordinator)}}
\INDENT{\textbf{upon} receiving (\textsc{Phase 1b}, \mv{rnd}, \mv{v-rnd}, \mv{v-val}) from $Q_a$
		such that $\mv{c-rnd} = \mv{rnd}$}
%	\STATE // $Q_a$ is a quorum with $\lceil N_a/2 \rceil$ acceptors
	\STATE let $k$ be the largest \mv{v-rnd} value received
	\STATE let $V$ be the set of (\mv{v-rnd},\mv{v-val}) received with $\mv{v-rnd}\!=\!k$
%	\IF{$k=0$}
%		\STATE let $\mv{c-val}$ be $v$
%	\ELSE
%		\STATE let $\mv{c-val}$ be the only element in $V$
%	\ENDIF
	\STATE \textbf{if} $k=0$ \textbf{then} let $\mv{c-val}$ be $v$
	\STATE \textbf{else} let $\mv{c-val}$ be the only \mv{v-val} in $V$
	\STATE \textbf{for all} $q \in N_a$ \textbf{do} send ($q$, (\textsc{Phase 2a}, \mv{c-rnd}, \mv{c-val}))
\ENDINDENT
\vspace{1.5mm}
\STATE \emph{\underline{Task~4 (acceptor)}}
\INDENT{\textbf{upon} receiving (\textsc{Phase 2a}, \mv{c-rnd}, \mv{c-val}) from coordinator}
	\IF{$\mv{c-rnd} \geq \mv{rnd}$}
		\STATE let \mv{rnd} be \mv{c-rnd}
		\STATE let \mv{v-rnd} be \mv{c-rnd}
		\STATE let \mv{v-val} be \mv{c-val}
		\STATE send (coordinator, (\textsc{Phase 2b}, \mv{v-rnd}, \mv{v-val}))
	\ENDIF
\ENDINDENT
\vspace{1.5mm}
\STATE \emph{\underline{Task~5 (coordinator)}}
\INDENT{\textbf{upon} receiving (\textsc{Phase 2b}, \mv{v-rnd}, \mv{v-val}) from $Q_a$
		such that $\mv{c-rnd} = \mv{v-rnd}$}
%	\STATE // $Q_a$ is a quorum with $\lceil N_a/2 \rceil$ acceptors
% 	\IF{for all received messages: $\mv{v-rnd} = \mv{c-rnd}$}
		\STATE \textbf{for all} $q \in N_l$ \textbf{do} send ($q$, (\textsc{Decision}, \mv{v-val}))
%	\ENDIF
\ENDINDENT
\label{alg:ce}
\end{distribalgo}
%\line(1,0){240}

\end{algorithm}

%\pagebreak
The coordinator starts Phase~2 after receiving a reply from an m-quorum (Task~3). Before proposing a value in Phase~2, the coordinator checks whether some acceptor has already cast a vote in a previous round. This mechanism guarantees that only one value can be chosen in an instance of consensus. If an acceptor has voted for a value in a previous round, then the coordinator will propose this value; otherwise, if no acceptor has cast a vote in a previous round, then the coordinator can propose the value received from the proposer. In some cases it may happen that more than one acceptor have cast a vote in a previous round. In this case, the coordinator picks the value that was voted for in the highest-numbered round. From the algorithm, two acceptors cannot cast votes for different values in the same round.

An acceptor will vote for a value \mv{c-val} with corresponding round \mv{c-rnd} in Phase~2 if the acceptor has not received any Phase~1 message for a higher round (Task~4). Voting for a value means setting the acceptor's variables \mv{v-rnd} and \mv{v-val} to the values sent by the coordinator. If the acceptor votes for the value received, it replies to the coordinator. When the coordinator receives replies from an m-quorum (Task~5), it knows that a value has been decided and sends the decision to the learners.

In order to know whether their values has been decided, proposers are typically also learners. If a proposer does not learn its proposed value after a certain time (e.g., because its message to the coordinator was lost), it proposes the value again. As long as a nonfaulty coordinator is eventually selected, there is a majority quorum of nonfaulty acceptors, and at least one nonfaulty proposer, every consensus instance will eventually decide on a value.

%\begin{figure*}[h]
% \begin{center}
% 	\begin{tabular}{c@{}c}
%	\includegraphics[width=0.5\columnwidth]{figures/line2.pdf}&
%	\includegraphics[width=0.5\columnwidth]{figures/star-1.pdf}\\
%	\end{tabular}      
%\caption{}
%\label{fig:obs1}
%\end{center}
%\end{figure*}

Algorithm~1 can be optimized in a number of ways~\cite{Lam98}. The coordinator can execute Phase~1 before a value is received from a proposer. In doing so, once the coordinator receives a value from a proposer, consensus can be reached in four communication steps, as opposed to six. Moreover, if acceptors send Phase~2B messages directly to the learners, the number of communication steps for a decision is further reduced to three (see Figure~\ref{fig:archh1}(a)). 

\section{Ring Paxos}
\label{sec:rpaxos}

Our Ring Paxos protocols are variations of Paxos, optimized for clustered systems. 
In Section~\ref{sec:perfcons} we motivate the most important design decisions behind our protocols.
In Sections~\ref{sec:multirpaxos} and~\ref{sec:unirpaxos} we detail the multicast-based and unicast-based versions of Ring Paxos under normal conditions, that is, in the absence of process crashes and message losses. 
In Sections~\ref{sec:handlingfail}, \ref{sec:flowcontrol}, and~\ref{sec:gc} we discuss, respectively, abnormal conditions, flow control, and garbage collection in Ring Paxos. We argue for the correctness of the protocols in the Appendix.

\subsection{Design considerations}
\label{sec:perfcons}
Ring Paxos was motivated by best practices in the design of networked systems where messages are subject to small delays (e.g., a local-area network).\footnote{While Ring Paxos could be deployed in a wide-area environment, in such settings latency, not throughput, tends to be the parameter of optimization.}
We designed Ring Paxos around two main ideas: the separation of message ordering and payload propagation, and  efficient means to carry out these two tasks.

Atomic broadcast protocols typically require several rounds of message exchange, separating message ordering from payload propagation saves bandwidth: the payload needs to be reliably propagated to the acceptors and learners once.  Care must be taken to ensure that at least $f+1$ acceptors store the message payload before it is ordered however, otherwise we risk losing payloads.  Ring Paxos solves this issue with a simple technique, as we explain next.

To efficiently order messages, Phase 2 of Ring Paxos relies on a communication pattern called {\it pipelined communication}. 
Acceptors are placed along a logical ring and when an acceptor receives a Phase 2 message (either a Phase 2A or 2B), it forwards this message along with its own Phase 2B message to the next acceptor in the ring.
This process continues until the coordinator is reached.
Pipelined communication allows extensive use of batching, which results in fewer protocol messages: Phase 2 messages tend to be small (recall that we only order message identifiers) and many of these messages can be batched together.
In addition, this communication scheme better balances the incoming and outgoing communication at the acceptors and coordinator.

\begin{figure*}[ht]
 \begin{center}
\begin{tabular}{c@{}c}
	 \includegraphics[width=\columnwidth]{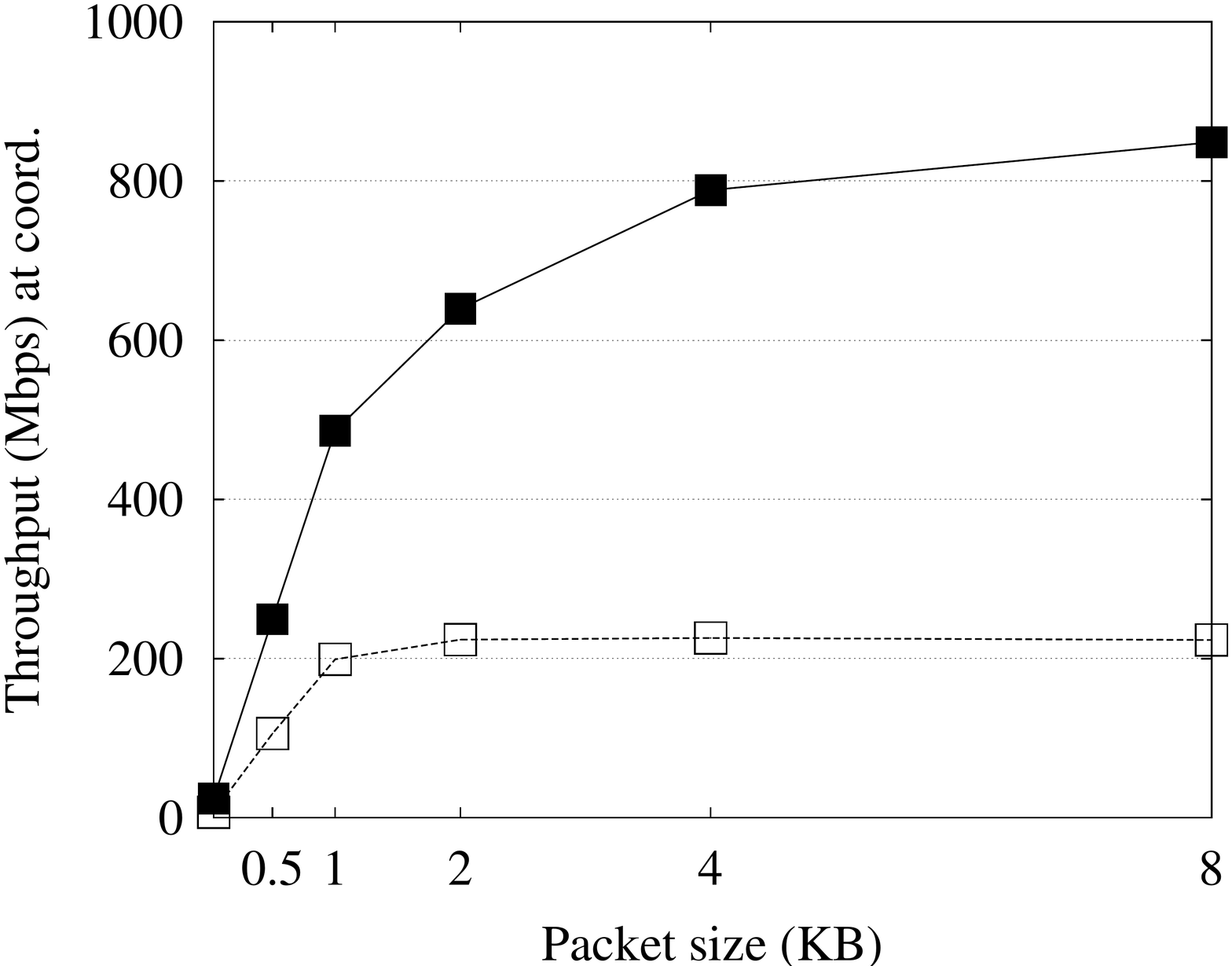} &
	 \includegraphics[width=\columnwidth]{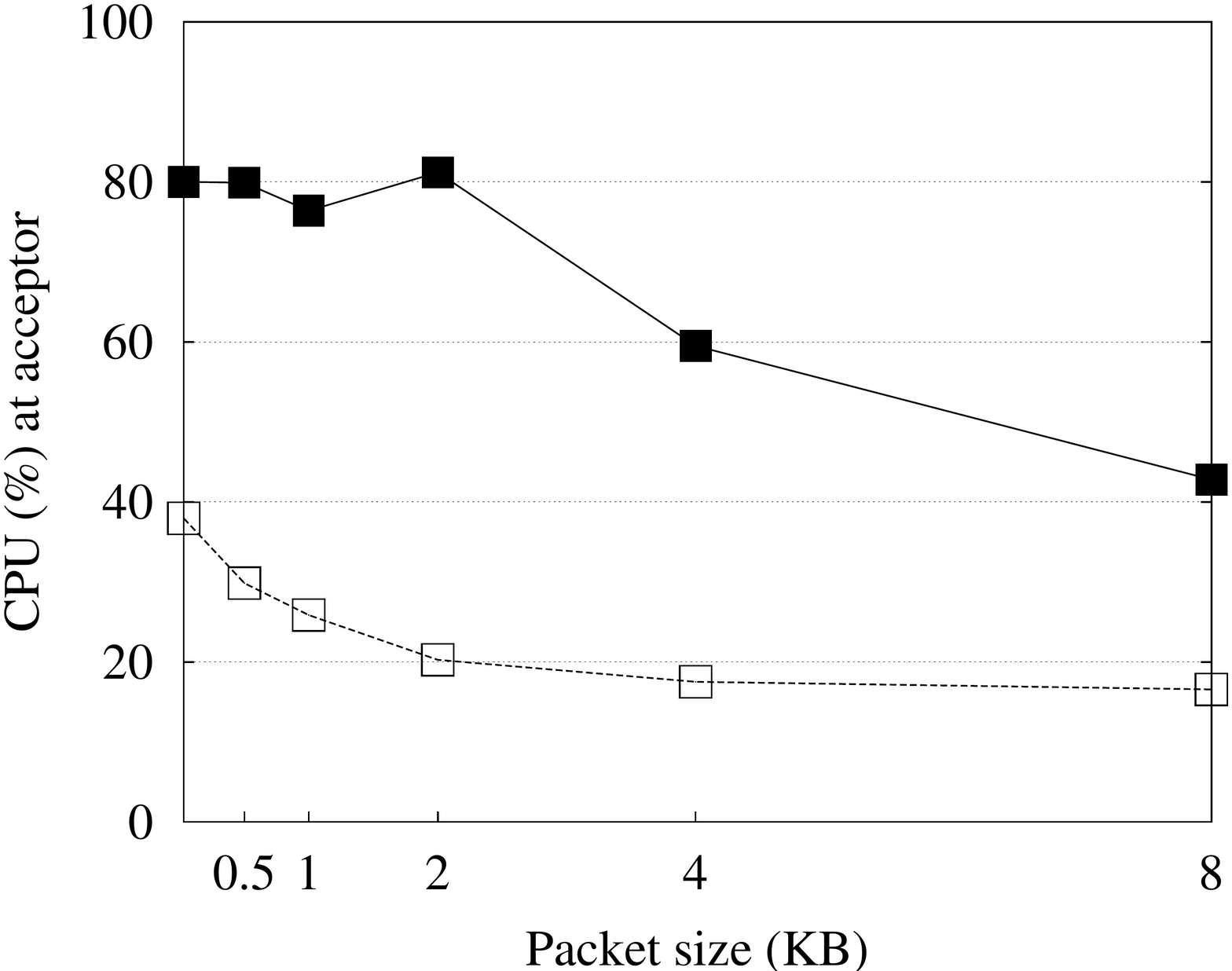}\\
           \includegraphics[width=\columnwidth]{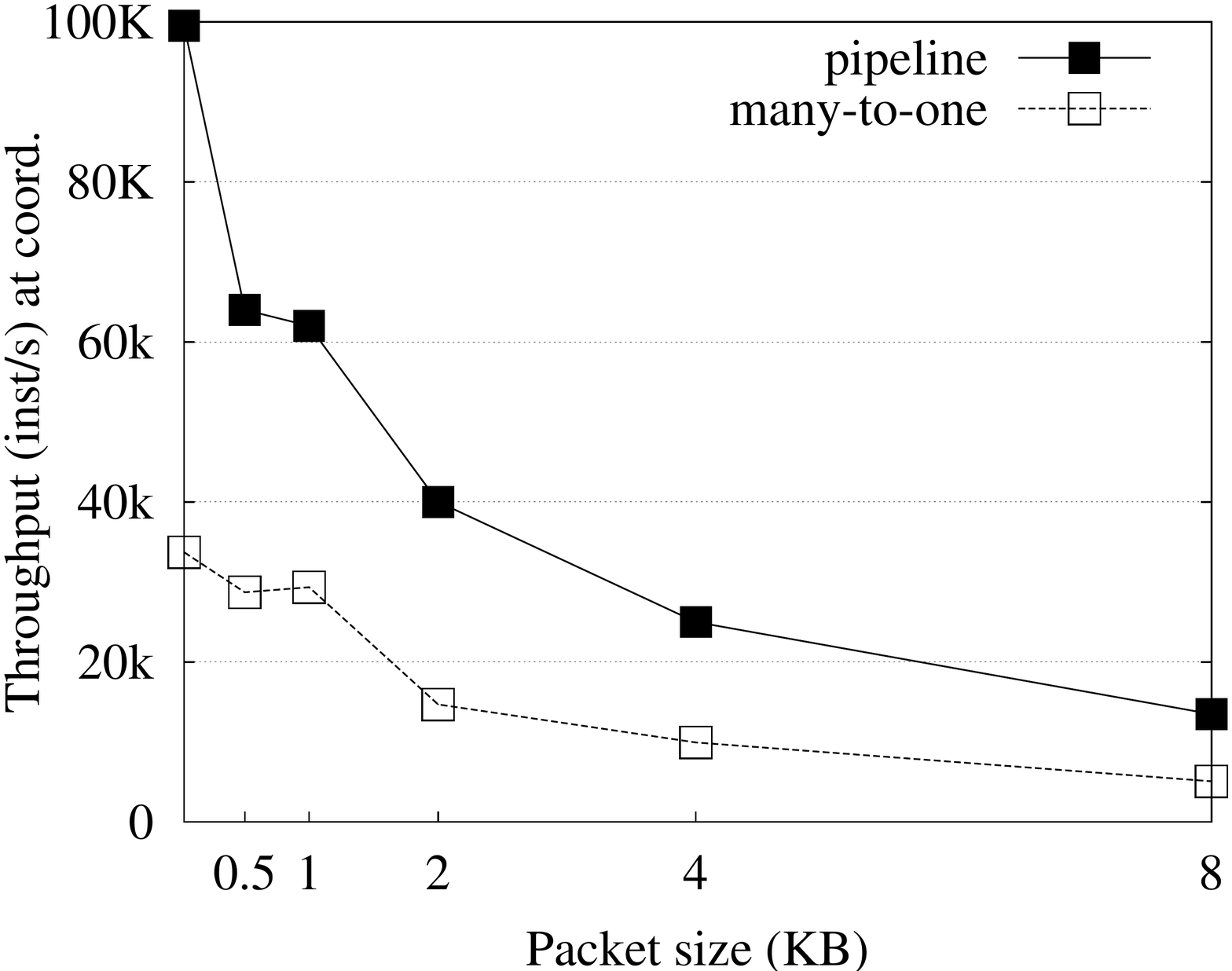} &
           \includegraphics[width=\columnwidth]{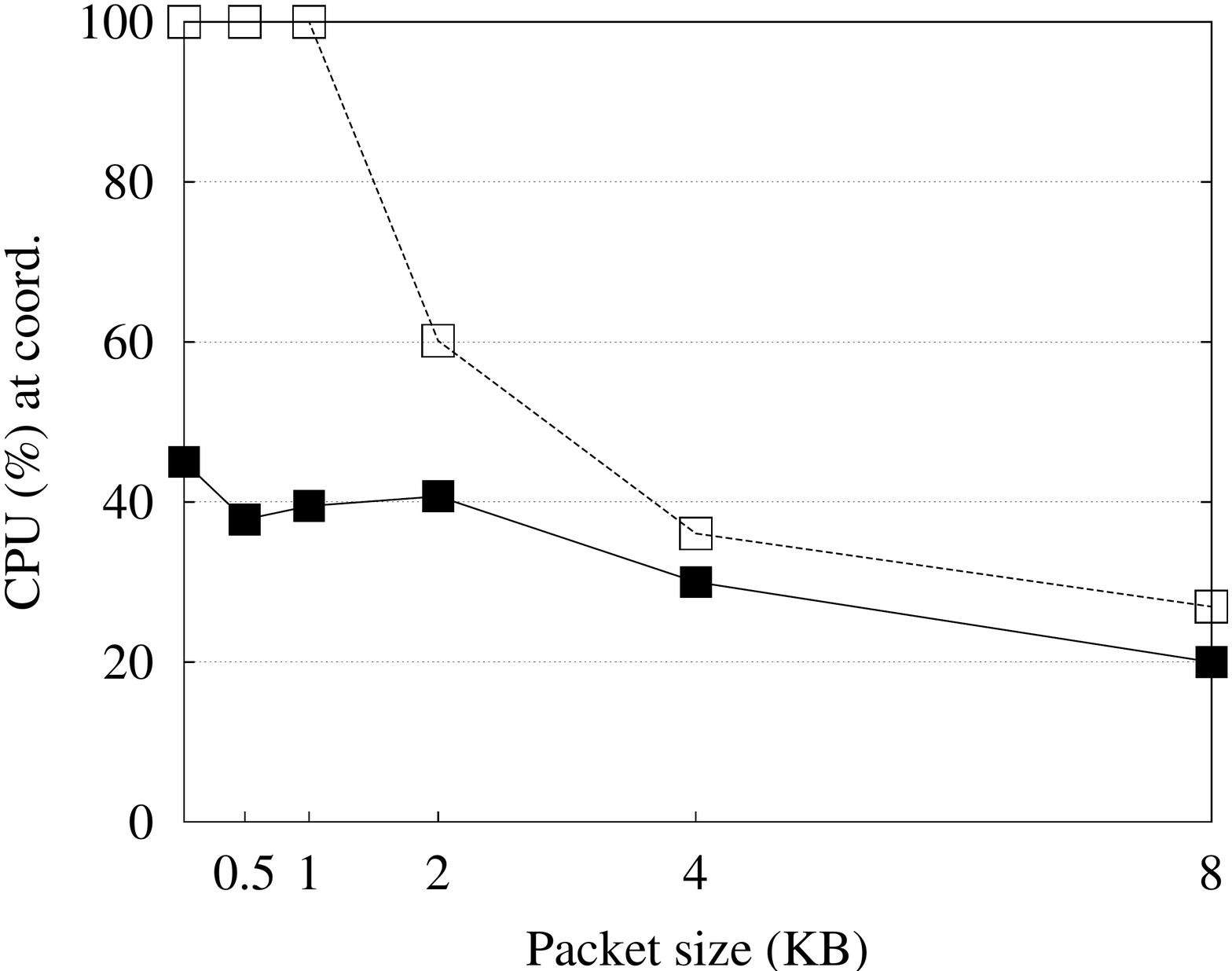} \\
\end{tabular}      
\caption{How to efficiently propagate Phase~2B messages to the coordinator with four acceptors and one coordinator. The throughput in the top left graph is the receiving throughput from only one of the incoming links of the coordinator. The number of incoming links at coordinator for the pipeline and many-to-one patterns is one and four respectively. As an example in this graph when the throughput shown for many-to-one is 200 Mbps the aggregate incoming bandwidth consumed is 800 Mbps. (The left-most value in all the graphs corresponds to packets with 32 bytes.) }
\label{fig:obs1}
\end{center}
%\vspace{-20mm}
\end{figure*}

\begin{figure*}
 \begin{center}
  	\begin{tabular}{c@{}c}
      \includegraphics[width=\columnwidth]{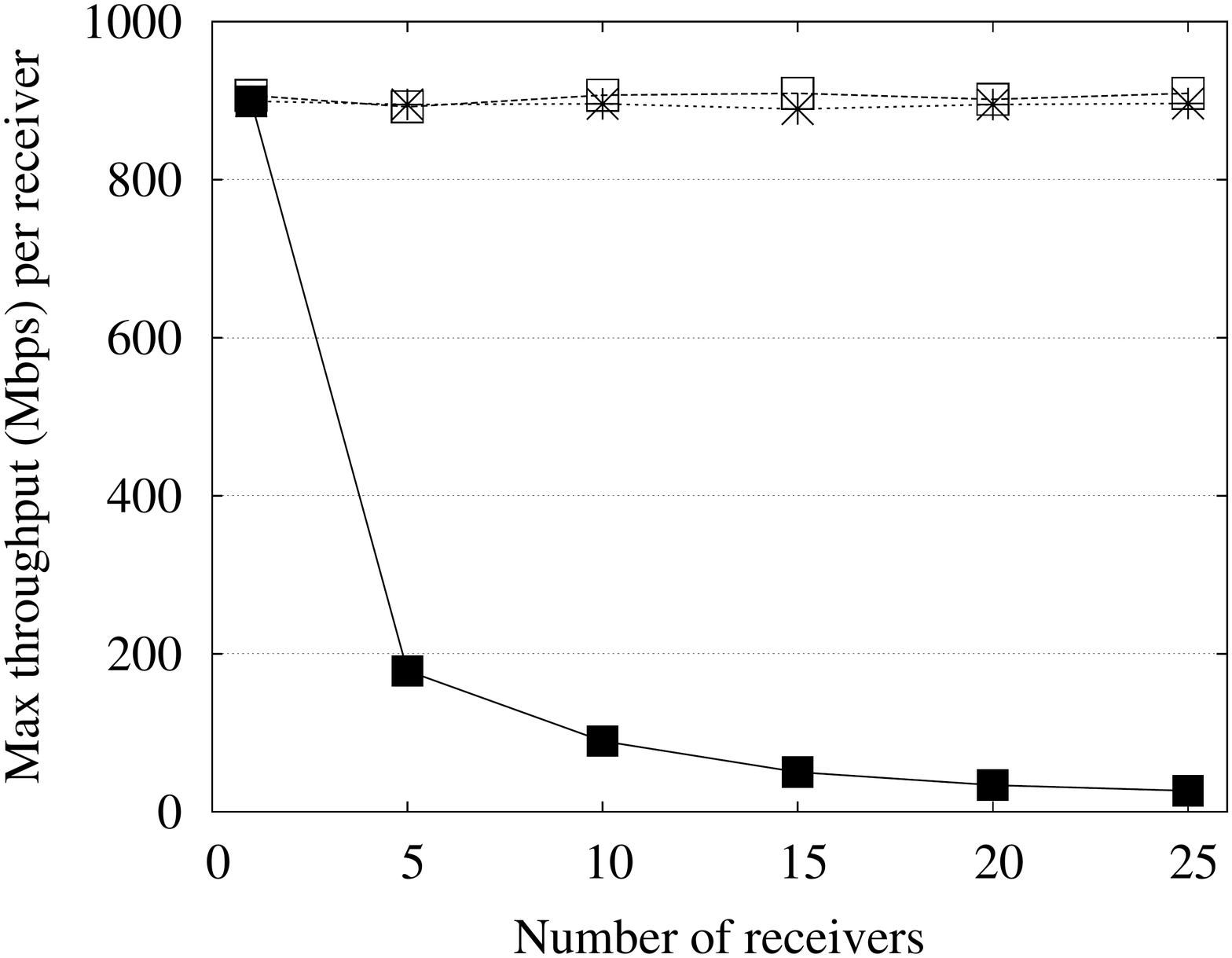} &
       \includegraphics[width=\columnwidth]{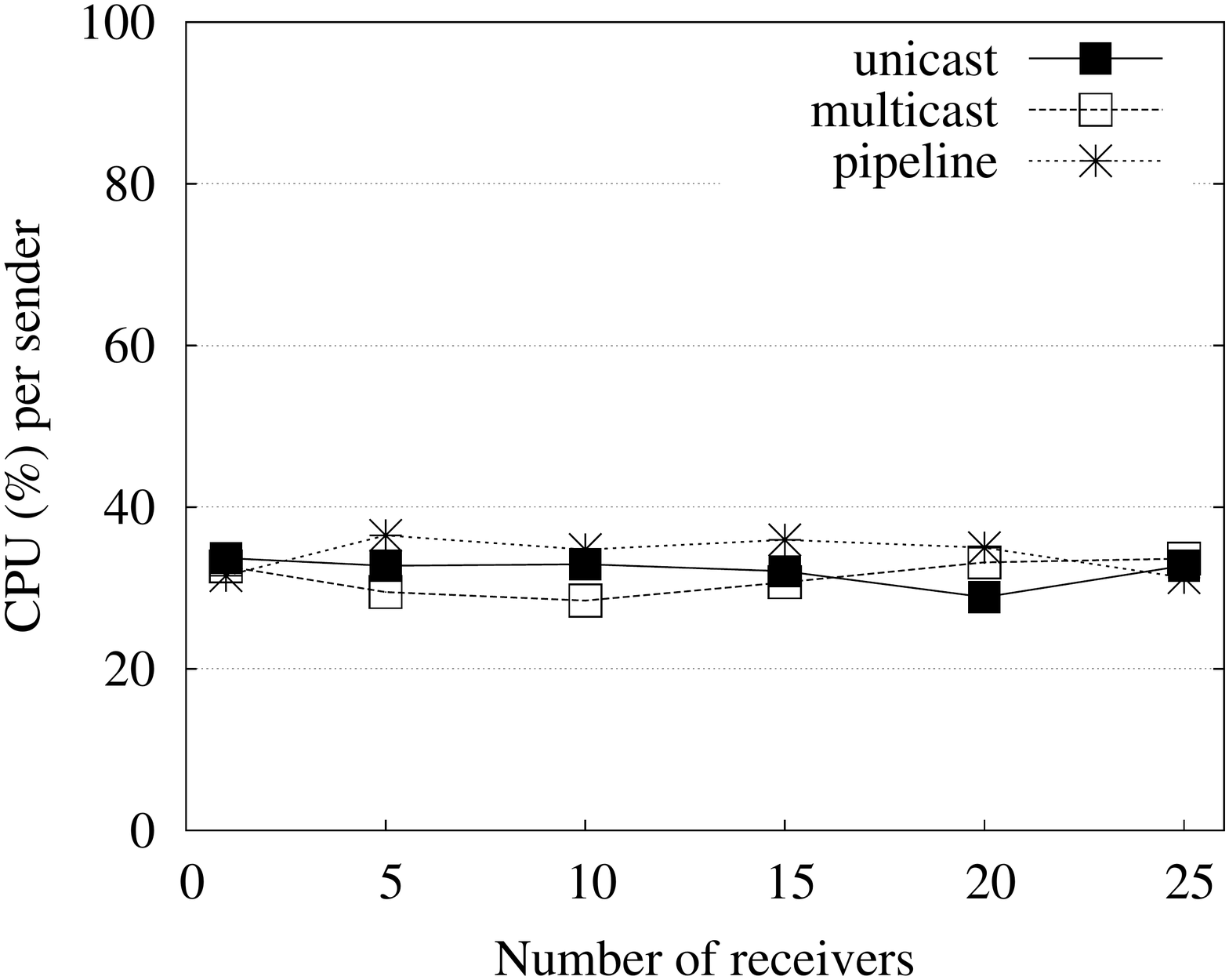} \\
       \end{tabular}
     \caption{Performance comparison of unicast, multicast and pipeline to propagate payloads.}
     \label{fig:obs2}
\end{center}
\end{figure*}

We illustrate the advantage of pipelined communication by considering the standard {\it many-to-one communication} pattern, where upon receiving a Phase~2A message from the coordinator, an acceptor directly replies to the coordinator.
In pipelined communication, upon receiving a Phase~2A message, the first acceptor at the head of the pipeline builds a Phase~2B message and forwards it to its successor. 
This Phase~2B message contains an integer field whose value is incremented by each acceptor along the ring as an indication of the acceptor's vote, without additional Phase~2B messages being initiated on the path to the coordinator. 
Thus, the size of a Phase~2B message that reaches the coordinator in the pipelined communication is equal to the size of individual Phase~2B messages that acceptors send to the coordinator in the many-to-one communication pattern. 
Therefore, to make one decision the coordinator consumes $n$ times more incoming bandwidth and CPU in the many-to-one communication pattern than in pipelined communication pattern, where $n$ is the number of acceptors other than the coordinator. 

We compare the efficiency of pipelined and many-to-one communication in an experiment with five acceptors, where one acceptor plays the role of the coordinator. The top left graph of Figure~\ref{fig:obs1} shows the incoming throughput at the coordinator from only one of the acceptors connected to it. In pipeline there is only one acceptor directly connected to the coordinator whereas in the many-to-one there are four. 
The bottom left graph shows the corresponding number of instances decided at the coordinator. 
The graphs on the right show the CPU usage at the acceptors (top) and the coordinator (bottom). 
For the pipelined communication the CPU is shown for an acceptor at the middle of the path. 
Pipeline is more CPU intensive at the acceptors as each acceptor (except the first one) both receives and forwards a Phase~2B message, whereas the coordinator only receives Phase~2B messages. 
However, as we will show later, acceptors are not the most CPU-loaded nodes in Ring Paxos, when the rest of the protocol is also considered. 
From the results, pipeline is preferable to many-to-one with respect to throughput. 
For small messages, this happens because pipeline is less CPU intensive at the coordinator. For large messages, the advantage stems from a balanced use of the incoming and outgoing bandwidth of the nodes.\footnote{For more details about these experiments, we refer the reader to Section~\ref{sec:perf}.}

The ring topology allows Ring Paxos to order message identifiers efficiently.  To obtain a high-throughput atomic broadcast protocol, we need a way to efficiently propagate payloads.  We do so by either relying on network-level multicast or the ring itself, in case multicast is not available.  Network-level multicast enables high-throughput propagation of messages to the nodes of a cluster~\cite{Vigfusson2010}. There are two reasons for this. First, multicast delegates to the interconnect (i.e., ethernet switch) the work of transferring messages to each one of the destinations. Second, to propagate a message to all destinations there is only one system call and one context switch from a user process to the operating system at the sender. If the sender contacts each destination individually using unicast communication, however, there is one system call and one context switch per destination. A throughput-efficient alternative to multicast is to have nodes communicate in a pipelined pattern such that each node sends the message only once, making better use of its CPU and bandwidth resources~\cite{Guerraoui2010}.

Figure~\ref{fig:obs2} assesses the performance of the three communication strategies. Clearly, multicast and pipeline unicast perform much better than one-to-many unicast (i.e., when the sender contacts each destination individually). Moreover, the results depicted in the graph on the bottom of Figure~\ref{fig:obs2} show that this performance advantage does not come at the expense of increased CPU utilization, as all three strategies compare similarly with respect to this metric. This happens because in all cases the nodes sending messages communicate at maximum rate.

\subsection{Multicast-based Ring Paxos (M-Ring Paxos)}
\label{sec:multirpaxos}

We now present the Ring Paxos algorithm based on multicast communication. In Algorithm~2, statements in gray are the same for Paxos and M-Ring Paxos. As in Paxos, the execution is divided in two phases. Moreover, the mechanism to ensure that only one value can be decided in an instance of consensus is the same as in Paxos.

\begin{figure*}[hbt]
    %\begin{center}
	 \includegraphics[width=1\textwidth]{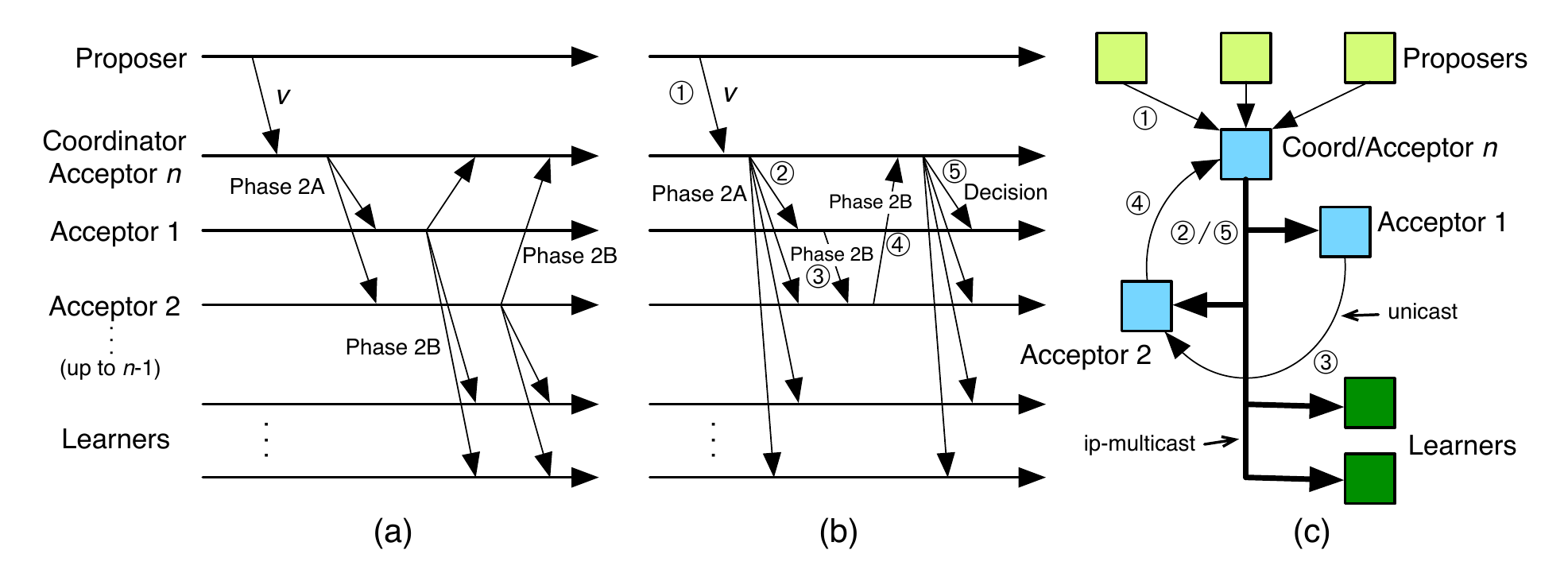}
	\vspace{-5mm}\caption{Optimized Paxos (a) and M-Ring Paxos (b,c).}
	\label{fig:archh1}
%	\vspace{-5mm}
   %\end{center}
\end{figure*}

Differently than Paxos, M-Ring Paxos disposes a majority quorum of acceptors (m-quorum) in a \emph{logical directed ring} (see Figure~\ref{fig:archh1}(b)(c)). The coordinator also plays the role of acceptor in M-Ring Paxos, and it is the last acceptor in the ring. Placing acceptors in a ring reduces the number of incoming messages at the coordinator and balances the communication among acceptors. Before the coordinator starts Phase~1, it proposes the ring on which acceptors will be located. By replying to the coordinator, the acceptors acknowledge that they abide by the proposed ring. Since Phase~1 can be executed before values are proposed, there is little gain in optimizing the sending of Phase 1B messages. Thus, we use the ring only to propagate Phase 2B messages. In addition to checking what value can be proposed in Phase~2 (Task~3), the coordinator also creates a unique identifier for the value to be proposed. Ring Paxos executes consensus on value ids~\cite{ESU04,LM04}; proposed values are disseminated to the m-quorum and to the learners in Phase~2A messages using ip-multicast.

\begin{algorithm}
%\line(1,0){240}

\begin{distribalgo}[1]
\small
\STATE \textbf{Algorithm 2: M-Ring Paxos} (for process $p$)
\vspace{1.5mm}
\STATE \emph{\underline{Task~1 (coordinator)}}
\INDENT{\bc{\textbf{upon} receiving value $v$ from proposer}}
	\STATE \bc{increase \mv{c-rnd} to an arbitrary unique value}
	%\STATE let \mv{c-ring} be an overlay ring with processes in $Q_a$
%	\STATE \textbf{for all} $q \in Q_a$ \textbf{do} send ($q$, (\textsc{Phase 1a}, \mv{c-rnd}, \mv{c-ring}))
	\STATE \textbf{for all} $q \in Q_a$ \textbf{do} send ($q$, (\textsc{Phase 1a}, \mv{c-rnd}))
\ENDINDENT
\vspace{1.5mm}
\STATE \emph{\underline{Task~2 (acceptor)}}
%\INDENT{\textbf{upon} receiving (\textsc{Phase 1a},\mv{c-rnd},\mv{c-ring}) from coordinator}
\INDENT{\textbf{upon} receiving (\textsc{Phase 1a},\mv{c-rnd}) from coordinator}
	\INDENT{\bc{\textbf{if} $\mv{c-rnd} > \mv{rnd}$ \textbf{then}}}
		\STATE \bc{let \mv{rnd} be \mv{c-rnd}}
%		\STATE let \mv{ring} be \mv{c-ring}
		\STATE \bc{send (coordinator, (\textsc{Phase 1b}, \mv{rnd}, \mv{v-rnd}, \mv{v-val}))}
	\ENDINDENT
\ENDINDENT
\vspace{1.5mm}
\STATE \emph{\underline{Task~3 (coordinator)}}
\INDENT{\textbf{upon} receiving (\textsc{Phase 1b}, \mv{rnd}, \mv{v-rnd}, \mv{v-val}) from $Q_a$ such that $\mv{rnd}=\mv{c-rnd}$}
	\STATE \bc{let $k$ be the largest \mv{v-rnd} value received}
	\STATE \bc{let $V$ be the set of (\mv{v-rnd},\mv{v-val}) received with $\mv{v-rnd}\!=\!k$}
	\STATE \bc{\textbf{if} $k=0$ \textbf{then} let $\mv{c-val}$ be $v$}
	\STATE \bc{\textbf{else} let $\mv{c-val}$ be the only \mv{v-val} in $V$}
	\STATE let \mv{c-vid} be a unique identifier for \mv{c-val}
	\STATE ip-multicast ($Q_a\!\cup\! N_l,$ (\textsc{Phase 2a}, \mv{c-rnd}, \mv{c-val}, \mv{c-vid}))
\ENDINDENT
\vspace{1.5mm}
\STATE \emph{\underline{Task~4 (acceptor)}}
\INDENT{\textbf{upon} ip-delivering (\textsc{Phase 2a}, \mv{c-rnd}, \mv{c-val}, \mv{c-vid})}
	\INDENT{\bc{\textbf{if} $\mv{c-rnd} \geq \mv{rnd}$ \textbf{then}}}
		\STATE \bc{let \mv{rnd} be \mv{c-rnd}}
		\STATE \bc{let \mv{v-rnd} be \mv{c-rnd}}
		\STATE \bc{let \mv{v-val} be \mv{c-val}}
		\STATE let \mv{v-vid} be \mv{c-vid}
		\INDENT{\textbf{if} $p = first(ring)$ \textbf{then}}
			\STATE send ($successor(p, ring)$, (\textsc{Phase 2b}, \mv{c-rnd}, \mv{c-vid}))
		\ENDINDENT
	\ENDINDENT
\ENDINDENT
\vspace{1.5mm}
\STATE \emph{\underline{Task~5 (coordinator and acceptors)}}
\INDENT{\textbf{upon} \mbox{receiving (\textsc{Phase 2b},\mv{c-rnd},\mv{c-vid})}}
	\INDENT{\textbf{if} $\mv{v-vid} = \mv{c-vid}$ \textbf{then}}
		\INDENT{\textbf{if} $p \neq last(ring)$ \textbf{then}}
			\STATE send ($successor(p, ring)$, (\textsc{Phase 2b}, \mv{c-rnd}, \mv{c-vid}))
		\ENDINDENT
		\INDENT{\textbf{else}}
			\STATE ip-multicast ($Q_a \cup N_l,$ (\textsc{Decision}, \mv{c-vid}))
		\ENDINDENT
	\ENDINDENT
\ENDINDENT
\vspace{5mm}
\underline{Note:} \hspace{2mm}$first(ring)$: process that succeeds the coordinator in\\
\hspace{9mm} $ring$\\
\hspace{10mm}$last(ring)$: the coordinator process in $ring$\\
\hspace{10mm}$successor(p, ring)$: process that succeeds $p$ in $ring$
\label{alg:rpaxos}
\end{distribalgo}
%\line(1,0){240}
%\\

\end{algorithm}

Upon ip-delivering a Phase~2A message (Task~4), an acceptor checks that it can vote for the proposed value. If so, it updates its \mv{v-rnd} and \mv{v-val} variables, as in Paxos, and its \mv{v-vid} variable. Variable \mv{v-vid} contains the unique identifier of the proposed value; it is initialized with null. The first acceptor in the ring sends a Phase~2B message to its successor in the ring. Although learners also ip-deliver the proposed value, they do not learn it since it has not been accepted yet. The next acceptor in the ring to receive a Phase~2B message (Task~5) checks whether it has ip-delivered the value proposed by the coordinator in a Phase~2A message---the acceptor can only vote if it has received the value, not only its unique identifier. The check is done by comparing the acceptor's \mv{v-vid} variable to the value's identifier calculated by the coordinator---this is to ensure that when consensus is reached, a majority of acceptors knows the chosen value and this value can be retrieved by learners at any time. If the condition holds, then there are two possibilities: either the acceptor is not the coordinator (i.e., the last process in the ring), in which case it sends a Phase 2B message to its successor, or it is the coordinator and then it ip-multicasts a decision message including the identifier of the chosen value. Once a learner ip-delivers this message, it can learn the value received previously from the coordinator in the Phase 2A message.

Ring Paxos can make use of a number of optimizations, most of which have been described previously in the literature: when a new coordinator is elected, it executes Phase~1 for several consensus instances~\cite{Lam98}; Phase~2 is executed for a batch of proposed values, and not for a single value~(e.g., \cite{KA08}); one consensus instance can be started before the previous one has finished~\cite{Lam98}.

\begin{figure*}
  \begin{center}
	\includegraphics[width=\textwidth]{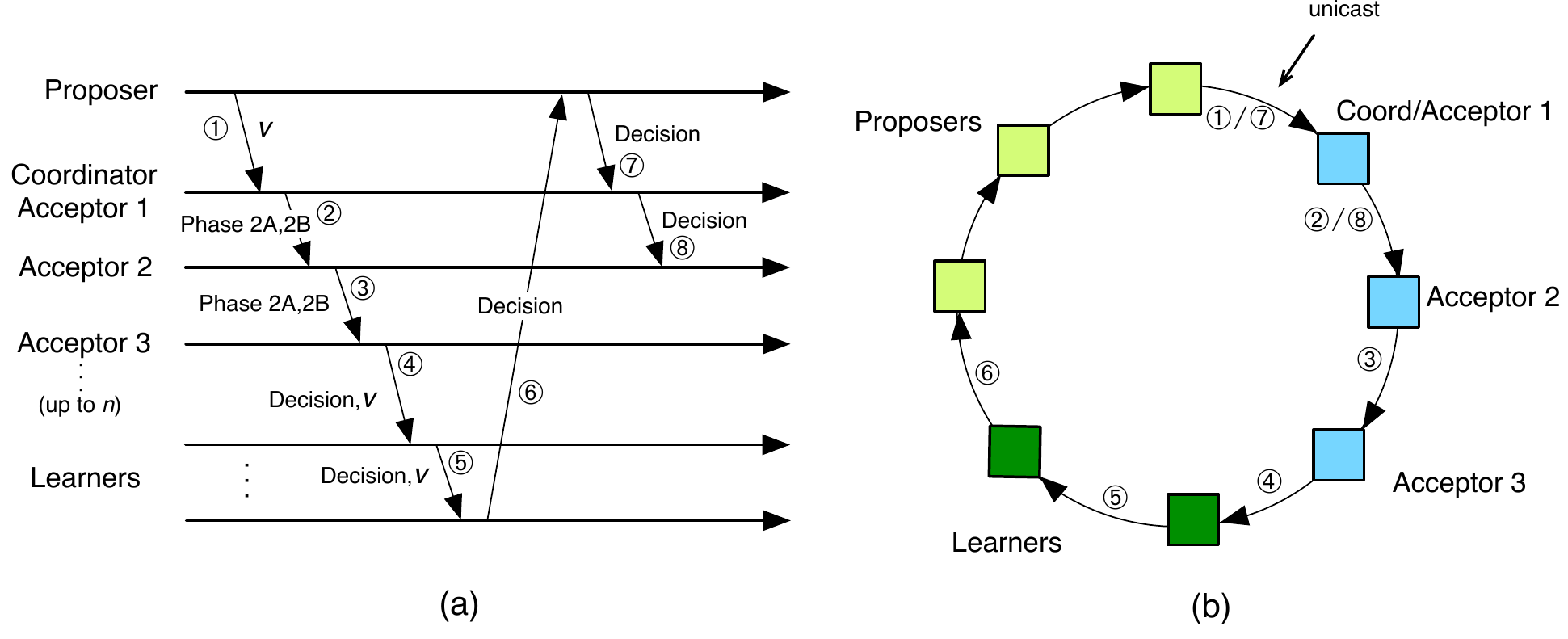}
	\vspace{-8mm}\caption{U-Ring Paxos.}
\label{fig:arch2}
  \end{center}
\end{figure*}

Placing an m-quorum in the ring, as opposed to placing all acceptors in the ring, reduces the number of communication steps to reach a decision. The remaining acceptors are spares, used only when an acceptor in the ring fails.\footnote{This idea is conceptually similar to Cheap Paxos~\cite{LM04}, although Cheap Paxos uses a reduced set of acceptors in order to save hardware resources, and not to reduce latency.} Finally, although ip-multicast is used by the coordinator in Tasks~3 and~5, this can be implemented more efficiently by overlapping consecutive consensus instances, such that the message sent by Task~5 of consensus instance $i$ is ip-multicast together with the message sent by Task~3 of consensus instance $i+1$.

\subsection{Unicast-based Ring Paxos (U-Ring Paxos)}
\label{sec:unirpaxos}

In the following, we present U-Ring Paxos. In Algorithm~3, statements in gray are the same for M-Ring Paxos and U-Ring Paxos. The execution is divided in two phases and the mechanism to ensure that only one value can be decided in an instance of consensus is the same as in Paxos.

U-Ring Paxos disposes proposers, learners and a majority-quorum of acceptors in a logical directed ring (see Figure~\ref{fig:arch2}). Pipelining all the processes in a ring is an alternative to multicast that can reach high throughput, a fact that was proved in~\cite{Guerraoui2010}. Processes in the ring can assume multiple roles and there is no restriction on the relative position of these processes in the ring, regardless of their roles. However, for simplicity of discussion, hereafter it is assumed that acceptors are lined up one after the other in the ring (see Figure~\ref{fig:arch2}(b)). To reduce latency, the coordinator is the first acceptor in the ring.

Once a proposer proposes a value, it is forwarded along the ring until it reaches the coordinator, which will proceed with Phase~1 as in Paxos. When the coordinator receives Phase~1B messages from an m-quorum (Task~3), it will check which value can be proposed and assign a unique identifier to the value to be proposed as in M-Ring Paxos. The coordinator then sends Phase~2A and Phase~2B messages to its successor in the ring (Task~3).
Similarly to Paxos and M-Ring Paxos, the coordinator in U-Ring Paxos can execute Phase 1 before a value is proposed, reducing the latency of the protocol.

Upon receiving a Phase~2A/2B message (Task~4), an acceptor checks that it can vote for the proposed value. If so, it updates its \mv{v-rnd}, \mv{v-val}, and \mv{v-vid} variables. If the acceptor does not precede the last acceptor in the ring, it sends the Phase~2A/2B message to its successor.
Differently than M-Ring Paxos where the coordinator is the one who checks whether a decision has been made in the instance, in U-Ring Paxos this is delegated to the last acceptor in the ring. After deciding, the last acceptor sends the decision, possibly together with the value chosen, to its successor in the ring.

Forwarding the chosen-value ends at the predecessor of the process who proposed the chosen value as at this point the value has been received by all the processes in the ring.
The decision, i.e., the chosen-value identifier, should be forwarded along the ring until it reaches the predecessor of the last acceptor (Task~5). 

\subsection{Handling abnormal cases}
\label{sec:handlingfail}

We now discuss how Ring Paxos tolerates lost messages and process crashes. The simplest way to handle abnormal cases is to switch back to original Paxos. A coordinator that suspects the failure of one or more acceptors may simply try to contact all acceptors in order to gather an m-quorum. This solution would reduce throughput but allow progress despite failures. 

Ring Paxos recovers from network and process failures as follows. Lost messages are handled with retransmissions. If the coordinator does not receive a response to its Phase~1A\,/\,2A messages, it re-sends them, possibly with a bigger round number. Eventually the coordinator will receive a response or will suspect the failure of a process (this suspicion might be erroneous). In this situation, the coordinator  lays out a new ring, excluding the suspected process, and re-executes Phase~1. 

With M-Ring Paxos, message loss may cause learners to receive only the value proposed and not the notification that it was accepted, only the notification without the value, or none of them. Learners can recover lost messages by inquiring other processes. M-Ring Paxos assigns each learner to a preferential acceptor in the ring, which the learner can ask for lost messages. With U-Ring Paxos the value is received by a learner only after it was accepted. 

\begin{algorithm}
%\line(1,0){240}
\begin{distribalgo}[1]
\small
\STATE \textbf{Algorithm 3: U-Ring Paxos} (for process $p$)
\vspace{1.5mm}
%\STATE \emph{\underline{Initialization}}
%\STATE let \mv{ring} be an overlay ring with processes in $Q_a \cup N_{lp}$
%\vspace{1.5mm}
\STATE \emph{\underline{Task~1 (all)}}
\INDENT{\textbf{upon} receiving value $v$ proposed by \mv{P(v)} from $predecessor(p, ring)$}
	\IF{$p =$ coordinator}
		\STATE \bc{increase \mv{c-rnd} to an arbitrary unique value}
		%\STATE \bc{let \mv{c-ring} be an overlay ring with processes in $Q_a \cup N_{lp}$}
		%\STATE \bc{\textbf{for all} $q \in Q_a$ \textbf{do} send ($q$, (\textsc{Phase 1a}, \mv{c-rnd}, \mv{c-ring}))}
		\STATE \bc{\textbf{for all} $q \in Q_a$ \textbf{do} send ($q$, (\textsc{Phase 1a}, \mv{c-rnd}))}
	\ELSE
		\STATE send $v$ to $successor(p, ring)$
	\ENDIF
\ENDINDENT
\vspace{1.5mm}
\STATE \emph{\underline{Task~2 (acceptor)}}
%\INDENT{\bc{\textbf{upon} receiving (\textsc{Phase 1a},\mv{c-rnd},\mv{c-ring}) from coordinator}}
\INDENT{\bc{\textbf{upon} receiving (\textsc{Phase 1a},\mv{c-rnd}) from coordinator}}
	\INDENT {\bc{\textbf{if} $\mv{c-rnd} > \mv{rnd}$ \textbf{then}}}
		\STATE \bc{let \mv{rnd} be \mv{c-rnd}}
%		\STATE \bc{let \mv{ring} be \mv{c-ring}}
		\STATE \bc{send (coordinator, (\textsc{Phase 1b}, \mv{rnd}, \mv{v-rnd}, \mv{v-val}))}
	\ENDINDENT
\ENDINDENT
\vspace{1.5mm}
\STATE \emph{\underline{Task~3 (coordinator)}}
\INDENT{\bc{\textbf{upon} receiving (\textsc{Phase 1b}, \mv{rnd}, \mv{v-rnd}, \mv{v-val}) from $Q_a$ such that $\mv{rnd}=\mv{c-rnd}$}}
	\STATE \bc{let $k$ be the largest \mv{v-rnd} value received}
	\STATE \bc{let $V$ be the set of (\mv{v-rnd},\mv{v-val}) received with $\mv{v-rnd}\!=\!k$}
	\STATE \bc{\textbf{if} $k=0$ \textbf{then} let $\mv{c-val}$ be $v$}
	\STATE \bc{\textbf{else} let $\mv{c-val}$ be the only \mv{v-val} in $V$}
	\STATE \bc{let \mv{c-vid} be a unique identifier for \mv{c-val}}
	\STATE {send ($successor(p, ring)$, (\textsc{Phase 2a/2b}, \mv{c-rnd}, \mv{c-val}, \mv{c-vid}))}
\ENDINDENT
%	\INDENT {\textbf {if} \mv{successor} $!=$ \mv{P(c-val)}}
%		\STATE send (successor, (\textsc{Phase 2a, Phase 2b}, \mv{c-rnd}, \mv{c-val}, \mv{c-vid}))
%	\ENDINDENT
%	\INDENT {\textbf{else}}
%		\STATE send (successor, (\textsc{Phase 2b}, \mv{c-rnd}, \mv{c-vid}))
%	\ENDINDENT

\vspace{1.5mm}
\STATE \emph{\underline{Task~4 (acceptor)}}
\INDENT{\textbf{upon} receiving (\textsc{Phase 2a/2b}, \mv{c-rnd}, \mv{c-val}, \mv{c-vid})}
	\INDENT{\bc{\textbf{if} $\mv{c-rnd} \geq \mv{rnd}$ \textbf{then}}}
	%\IF{$\mv{c-rnd} \geq \mv{rnd}$}
		\STATE \bc{let \mv{rnd} be \mv{c-rnd}}
		\STATE \bc{let \mv{v-rnd} be \mv{c-rnd}}
		\STATE \bc{let \mv{v-val} be \mv{c-val}}
		\STATE \bc{let \mv{v-vid} be \mv{c-vid}}
		\IF{$p = last\_acceptor(ring)$}
			\STATE \mv{Send\_Decision}(\mv{c-vid}, \mv{c-val})
		\ELSE
			\STATE send ($successor(p, ring)$, (\textsc{Phase 2a/2b}, \mv{c-rnd}, \mv{c-val}, \mv{c-vid}))
		\ENDIF
	%\ENDIF
	\ENDINDENT
\ENDINDENT

\vspace{1.5mm}
\STATE \emph{\underline{Task~5 (all)}}
\INDENT{\textbf{upon} \mbox{receiving (\textsc{Decision},  \mv{c-vid}, \mv{c-val})}}
	\INDENT{\textbf{if} $p \neq \mv{predecessor}(last\_acceptor(ring))$}
		\STATE {\mv{Send\_Decision}(\mv{c-vid}, \mv{c-val})}
	\ENDINDENT
\ENDINDENT

\vspace{1.5mm}
\STATE \emph{\underline{\mv{Send\_Decision}(\mv{c-vid}, \mv{c-val})}}
\INDENT{\textbf{if} $p \neq \mv{predecessor(P(c-val), ring)}$}
		\STATE send ($successor(p, ring)$, (\textsc{Decision}, \mv{c-vid}, \mv{c-val}))
\ENDINDENT
\INDENT{\textbf{else}}
		\STATE send ($successor(p, ring)$, (\textsc{Decision}, \mv{c-vid}, --))
\ENDINDENT
\vspace{5mm}
\underline{Note:} \hspace{2mm}$P(v)$: proposer of value $v$\\
\hspace{10mm}$predecessor(p, ring)$: process that precedes $p$ in $ring$\\
\hspace{10mm}$successor(p, ring)$: process that succeeds $p$ in $ring$\\
\hspace{10mm}$last\_acceptor(ring)$: the $f$-th acceptor after the\\
\hspace{10mm}coordinator in $ring$
\label{alg:rpaxos2}
\end{distribalgo}
%\line(1,0){240}
%\\

\end{algorithm}

Thus, the problem described above cannot arise. 
However, in both protocols learners may not have sufficient time to handle the decisions, and this may lead learners to drop some decisions. In fact, both versions of Ring Paxos need flow control  to mitigate this phenomenon. 
We discuss flow control in Section~\ref{sec:flowcontrol}.

With both protocols, when an acceptor replies to a Phase~1A or to a Phase~2A message, it must not forget its state (i.e., variables \mv{rnd}, \mv{ring}, \mv{v-rnd}, \mv{v-val}, and \mv{v-vid}) despite failures. There are two ways to ensure this. First, by assuming that a majority of acceptors never fails. Second, by requiring acceptors to keep their state on stable storage before replying to Phases~1A and~2A messages. 

Finally, a failed coordinator is detected by the other processes, which select a new coordinator. Before GST (see Section~\ref{sec:model}) it is possible that multiple coordinators co-exist. However, as Paxos, Ring Paxos guarantees safety even when multiple coordinators execute at the same time, although it may not guarantee liveness. After GST, eventually a single correct coordinator is selected.

\subsection{Flow control}
\label{sec:flowcontrol}

In Ring Paxos, flow control helps regulate the speed at which consensus instances are executed. In doing so, we not only reduce the likelihood of message loss for both unicast and multicast communications, but we also ensure that learners are given enough time to process decisions. For instance, when Ring Paxos is used to implement state-machine replication, decisions represent commands that read or write the application state, and may need extra processing time.

To illustrate why flow control is important, consider a scenario where commands are ordered faster than they can be applied at the leaners. Without flow control, buffers storing decisions to be processed will eventually overflow and learners will start dropping decisions, leading learners to do extra work to retrieve the lost decisions.  As a consequence, the performance of the replicated system may decrease, to the point where clients time out and retransmit their commands, leading to further inefficiencies.

Flow control in U-Ring Paxos is easily implemented. Communication between two consecutive processes in the ring is done using TCP and, on each process, TCP buffers are made sufficiently large to take into account the processing time of Phase 1 and 2 messages. To ensure that learners have sufficient time to handle the decided values, U-Ring Paxos (a)~lets learners apply a decision before forwarding it to the next process in the ring and (b)~limits the number of outstanding consensus instances.

In M-Ring Paxos, communication is based on UDP and flow control at the coordinator ensures that messages are sent at a rate the network can handle. With M-Ring Paxos, since learners are not part of the ring, we use the following mechanism for flow control. Learners constantly monitor their buffer for decisions that remain to be processed and when the used slots reach a threshold, they  notify one of the acceptors. The notification informs the acceptors about the number of unprocessed requests at the learner. Acceptors forward this notification along the ring until it reaches the coordinator. At this point, the coordinator reduces the \emph{window} of outstanding consensus instances accordingly, leading it to start fewer instances in parallel. Provided that the coordinator slows down sufficiently, learners will be able to apply decisions as fast as they are ordered,  and they will stop sending notifications to the acceptors. 

This technique allows learners to slow down the coordinator before decisions are dropped. In case some decisions do get lost, for instance if the coordinator starts with a window that is too large, lost decisions are retrieved from the acceptors. 

To allow M-Ring Paxos to recover from temporarily slow learners, the coordinator slowly increases the window size when no notifications to slow down have been received for some time.

\subsection{Garbage Collection}
\label{sec:gc}

Acceptors store variables for each instance of Ring Paxos they participate in. The variables are the highest-round \mv{rnd} in which the acceptor executed a Phase 1 or 2, the highest-round \mv{v-rnd} in which it cast a vote in Phase 2, as well as the corresponding value \mv{v-val} and value identifier \mv{v-vid} the acceptor voted for.  Variable \mv{rnd} is shared across consensus instances and does not need to be garbage collected.  The other variables are discarded when $f+1$ learners have applied the corresponding decision to their application state.  To do so, each learner maintains its \emph{version}, the largest instance for which it applied the corresponding decision---learners apply decisions in instance order so if a learner has applied decision of instance $x$, it also has applied all decisions of instances lower than $x$.  In M-Ring Paxos, each learner periodically communicates its version to one of the acceptors (learners are assigned different acceptors to balance the associated load), and acceptors propagate this information along the ring.  Once an acceptor receives a version from $f+1$ learners, it computes the smallest version received and garbage collects variables for instances up to the smallest version.  In U-Ring Paxos, learners are part of the ring and directly forward their version to their successor.

The M-Ring Paxos and U-Ring Paxos coordinators also store state, namely the highest-round started \mv{c-rnd} and the value picked \mv{c-val} for a particular instance and round.  Variable \mv{c-rnd}, similarly to \mv{rnd}, is shared across instances and does not need to be garbage collected.  The value picked for a given instance and round can be discarded as soon as the coordinator receives the corresponding Phase 2b messages from its predecessor.

Acceptors also store the decisions sent by the coordinator in order to let learners retrieve decisions that they may not have received.  These decisions can be discarded similarly as with the acceptor variables.  If a learner requests a decision that has been garbage collected, the learner can be brought up to date by communicating with a learner with a sufficiently recent version, i.e., one that is larger than the instance of the decision missing at the learner.  Such a learner will always exist since we garbage collect a decision only after it is reflected in the state of $f+1$ learners.

\section{Related work}
\label{sec:rwork}

Several papers argued that Paxos is not an easy algorithm to implement~\cite{CGR07,KA08,RVR11}. Essentially, this is because Paxos is a subtle algorithm that leaves many non-trivial design decisions open. Besides providing insight into these matters, two of these papers present performance results of their Paxos implementations. In contrast to our Ring Paxos protocols, none of these papers present algorithmic modifications to Paxos. In~\cite{SS12}, an analytical analysis of the impact of several optimizations on the performance of the original Paxos algorithm is presented.  The authors also present extensive experimental results of their implementation in both LAN and WAN environments.

Several papers have built on Paxos to provide various abstractions such as a storage system~\cite{B2011}, a locking service~\cite{Bur06}, and a distributed database~\cite{megastore}.  Some of the optimizations presented in these papers, such as executing read-only operations at a single learner while ensuring linearizability or using a log to speed-up the execution of write requests are orthogonal to the design of Ring Paxos and could be used here as well.

\begin{table*}
\center
\caption{Comparison of atomic broadcast algorithms ($f$: number of tolerated failures).}
\vspace{4mm}
\label{table:algo_comparison}

%\resizebox{\columnwidth}
%{!}{
\begin{tabular}{|c|c|c|c|c|} \hline

Algorithm				& Class	& Communication	& Number of  	& Synchrony	\\
					&		& steps			& processes	& assumption	\\ \hline \hline
LCR~\cite{Guerraoui2010}& comm.	& $2f$			& $f+1$		& strong		\\
                              		& history	&         			&			&			\\ \hline
%Totem~\cite{amir1995totem} & privilege & $(4f+3)$ & $2f+1$ & weak
%\\ \hline
Ring+FD~\cite{ESU04}  & privilege & $(f^2 + 2f)$  & $f(f+1)+1$ &
weak \\ \hline
S-Paxos~\cite{spaxos}  & & $5$  & $2f+1$ & weak \\ \hline

M-Ring Paxos & --- & $(f+3)$ & $2f+1$ & weak\\ \hline
U-Ring Paxos & --- & $5f$ & $2f+1$ & weak\\ \hline
\end{tabular}
%}

\end{table*}

Paxos is not the only algorithm to implement atomic broadcast. 
Some protocols implement atomic broadcast through the virtual synchrony model introduced by the Isis system~\cite{BT87}. With virtual synchrony, processes are part of a group. They may join and leave the group at any time. When processes are suspected of crashing they are evicted from the group; virtual synchrony ensures that processes observe the same sequence of group memberships or \emph{views} and non-faulty members deliver the same set of messages in each view. Implementing such properties requires solving consensus. 

The literature on atomic broadcast is abundant. In~\cite{DUS04}, five classes of broadcast algorithms have been identified: fixed sequencer, moving sequencer, destination agreement, communication history-based, and privilege-based. Below, we review the five classes of atomic broadcast protocols.

In fixed sequencer algorithms~(e.g., \cite{BSS91,KT91}), broadcast messages are sent to a distinguished process, called the sequencer, who is responsible for ordering these messages. The role of sequencer is unique and only transferred to another process in case of failure of the current sequencer. In this class of algorithms, the sequencer may eventually become the system bottleneck. 

Moving sequencer protocols are based on the observation that rotating the role of the sequencer distributes the load associated with ordering messages among processes. The ability to order messages is passed from process to process using a \emph{token}. The majority of moving sequencer algorithms are optimizations of~\cite{CM84}. These protocols differ in the way the token circulates in the system: in some protocols the token is propagated along a ring~\cite{CM84,CM95}, in others, the token is passed to the least loaded process~\cite{KK97}. All the moving sequencer protocols we are aware of are based on the broadcast-broadcast communication pattern. According to this pattern, to atomically broadcast a message $m$, $m$ is broadcast to all processes in the system; the token holder process then replies by broadcasting a unique global sequence number for $m$. High-throughput can be obtained by resorting to network-level broadcast.
%However, allowing multiple processes to broadcast at the same time will lead to message loss, which hurts performance.
%
Mencius is another moving sequencer-based protocol that implements state-machine replication and is derived from Paxos~\cite{mao2008mencius}. 
Mencius is designed for wide-area networks in which optimizing for latency is the main objective in contrast to throughput, the focus of Ring Paxos. 

Protocols falling in the destination agreement class compute the message order in a distributed fashion (e.g., \cite{CT96,FIMR98}). These protocols typically exchange a quadratic number of messages for each message broadcast, and thus are not good candidates for high throughput.

In communication history-based algorithms, the message ordering is determined by the message sender, that is, the process that broadcasts the message (e.g., \cite{Lam78b,Ng91}). Message ordering is usually provided using logical or physical time. Of special interest is LCR, which arranges processes along a ring and uses vector clocks for message ordering~\cite{Guerraoui2010}. This protocol has similar throughput to our Ring Paxos protocols but requires \emph{perfect failure detection}: erroneously suspecting a process to have crashed is not tolerated. Perfect failure detection implies strong synchrony assumptions about processing and message transmission times.

The last class of atomic broadcast algorithms, denoted as privilege-based, allows a single process to broadcast messages at a
time; the message order is thus defined by the broadcaster. Similarly to moving sequencer algorithms, the privilege to order messages circulates among broadcasters in the form of a token. Differently from moving sequencer algorithms, message ordering is provided by the broadcasters and not by the sequencers. In~\cite{amir1995totem}, the authors propose Totem, a protocol based on the virtual synchrony model.  In the case of process or network failures, the ring is reconstructed and the token regenerated using the new group membership. In~\cite{ESU04}, fault-tolerance is provided by relying on a failure detector; tolerating $f$ process failures requires a quadratic number of processes. A general drawback of privilege-based protocols is their high latency: before a process $p$ can totally order a message $m$, $p$ must receive the token, which delays $m$'s delivery.

M-Ring Paxos and U-Ring Paxos combine ideas from several broadcast protocols to provide high throughput and low latency. In this sense, they fit multiple classes, as defined above. To ensure high throughput, our protocols decouple message dissemination from ordering. The former is accomplished using ip-multicast or pipelined unicast; the latter is done using consensus on
message identifiers. To use the network efficiently, processes executing consensus communicate using a ring, similarly to the majority of privilege-based protocols. 

\begin{figure*}
  \begin{center}
    \begin{tabular}{c@{}c}
      \includegraphics[width=\columnwidth]{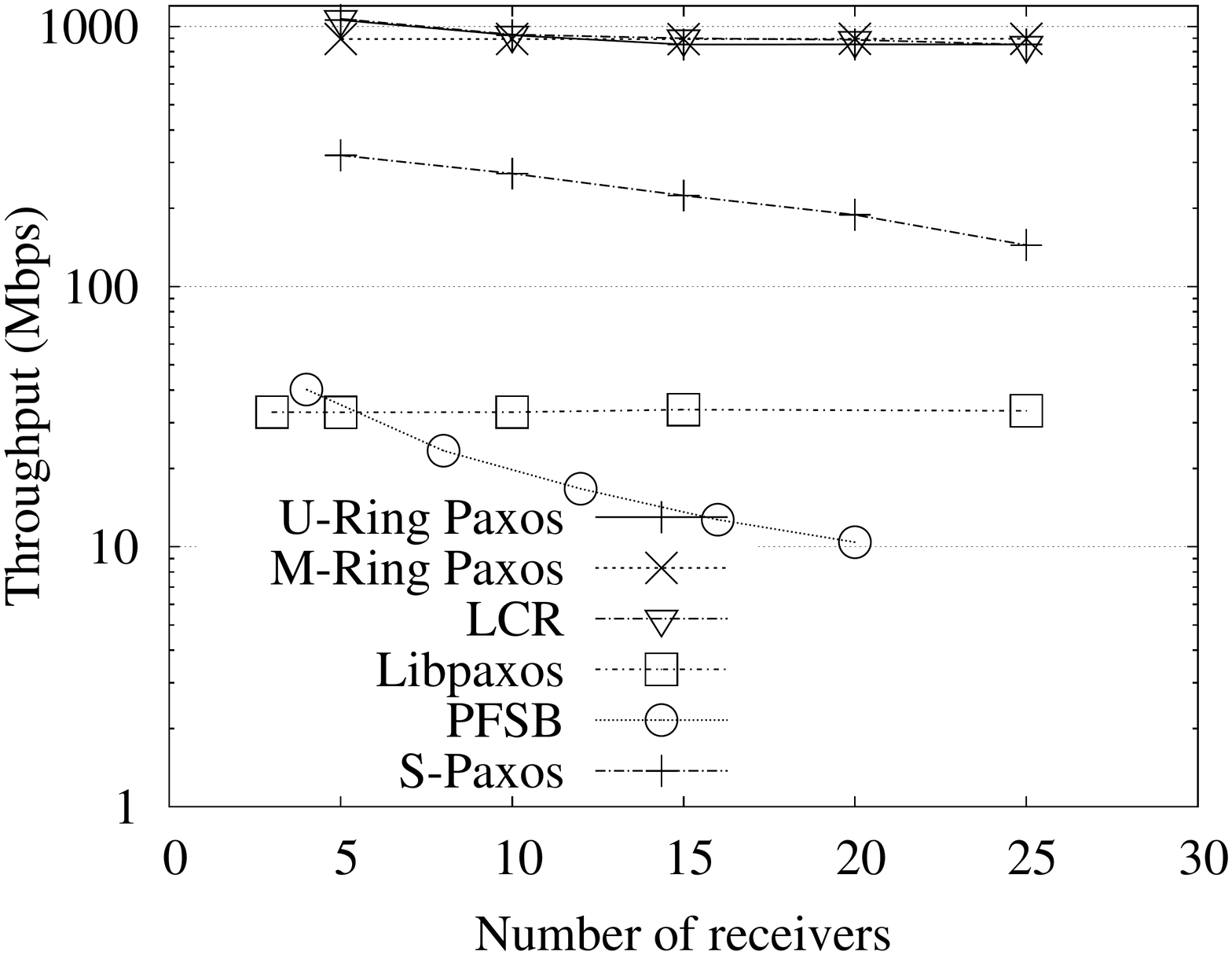} &
      \includegraphics[width=\columnwidth]{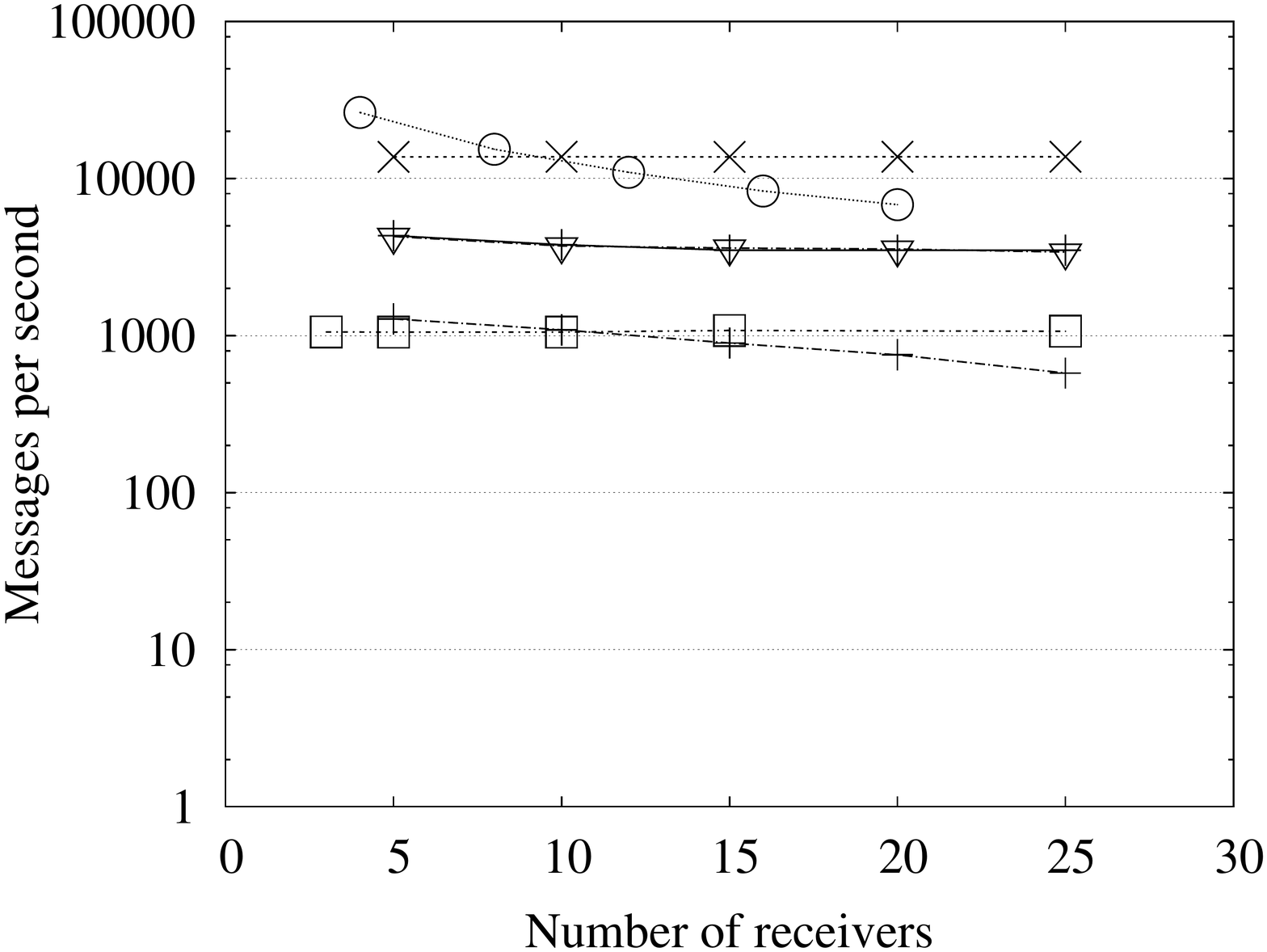} \\
    \end{tabular}
    \vspace{-3mm}
    \caption{Ring Paxos and other atomic broadcast protocols (message sizes c.f. Table~\ref{table:mte}). For Ring Paxos protocols $f$ is equal to two. In PFSB, U-Ring Paxos, and LCR the number of receivers is equal to the total number of processes. In Libpaxos and M-Ring Paxos it is equal to the number of learners.%, in Spread it is equal to the number of readers.
   }
  \label{fig:others}
  \end{center}
\end{figure*}

In Table~\ref{table:algo_comparison}, we compare algorithms that are closest to our Ring Paxos protocols in terms of throughput efficiency. All these protocols use a logical ring for process communication, which is a good communication pattern when optimizing for throughput. For each algorithm, we report its class, the minimum number of communication steps required by the last process to deliver a message, the number of processes required as a function of $f$, and the synchrony assumption needed for correctness. For the Ring Paxos protocols, we assume that each process plays the roles of proposer, acceptor, and learner. There are $f+1$ processes in the ring of M-Ring Paxos and $2f+1$ processes in the ring of U-Ring Paxos (i.e,. all processes are in the ring of U-Ring Paxos).

With M-Ring Paxos, delivery occurs as soon as messages make one revolution around the ring. Its latency is $f+3$ message delays since each message is first sent to the coordinator, circulates around the ring of $f+1$ processes, and is delivered after the final ip-multicast is received. With U-Ring Paxos, the worst case latency is $5f$. This happens when the process that broadcasts the message follows the coordinator in the ring. It takes $2f$ steps to reach the coordinator, and another $f$ steps for the decision. The decision must circulate around the ring in order to reach all processes, taking another $2f$ steps.

LCR requires two revolutions and thus has a latency in between the two Ring Paxos algorithms. In Totem, each message must also rotate twice along the ring to guarantee \emph{safe-delivery}, a property equivalent to uniform agreement: if a process (correct or 
not) delivers a message $m$ then all correct processes eventually deliver $m$. 
The atomic broadcast protocol in~\cite{ESU04} has a latency that is quadratic in $f$ since a ring requires more than $f^2$ nodes.

An efficient implementation of Paxos protocol is S-Paxos~\cite{spaxos}. The key idea in S-Paxos is to distribute the tasks of request reception and dissemination among all replicas. A client selects a replica arbitrarily and submits its requests to it. After receiving a request, a replica forwards it to all the other replicas. A replica receiving a forwarded request sends an acknowledgement to all other replicas. When a replica receives $f+1$ acknowledgements, it declares the request as stable. As in classical Paxos, the leader is responsible for ordering requests; differently from Paxos, ordering is performed on request ids. S-Paxos makes a balanced use of CPU and network resources; on the negative side, many messages must be exchanged before a request can be ordered. Due to the number of messages exchanged, this protocol is CPU-intensive.

\section{Performance evaluation}
\label{sec:perf}

In this section, we describe our Ring Paxos prototypes and the experimental evaluation we conducted. We consider the performance of Ring Paxos in the presence of message losses and in the absence of process failures. Process failures are hopefully rare events; message losses happen relatively often because of high network traffic.

We ran the experiments in a cluster of Dell SC1435 nodes equipped with 2 dual-core AMD-Opteron 2.0 GHz CPUs and 4GB of main memory. The servers were interconnected with an HP ProCurve2900-48G Gigabit switch (0.1 msec of round-trip time). For the experiments with disk writes we use OCZ-VERTEX3 SSDs. Each experiment (i.e., point in the graph) was repeated 3 to 10 times, with a few million messages broadcast in each execution. Every process is deployed on a dedicated node in the experiments.

In M-Ring Paxos, each process maintains a circular buffer of packets; each packet is 8 Kbytes long and the buffer is 160 Mbytes long. In U-Ring Paxos, processes maintain a circular buffer of packets, where each packet is 32 Kbytes long. U-Ring Paxos allocates 16 Mbytes of buffer space per proposer. For example, with five proposers the total space needed for buffers is 80 Mbytes.

\subsection{Implementation}

In M-Ring Paxos the acceptors and the learners use the buffer to match proposal ids to proposal contents, as these are decomposed by the coordinator. %In unicast-based Ring Paxos an additional level of matching proposer ids and proposal content is required. 
Messages received out of sequence (e.g., because of transmission losses) are stored in the buffer until they can be delivered (i.e., learned) in order.
Each packet sent by the coordinator is composed of two parts. In one part the coordinator stores the ids of decided values, and in the second part it stores new proposed values with their unique ids. 
%A buffer entry is freed after the coordinator has received the entry's corresponding Phase~2B message from its neighbor and ip-multicast a decision message related to the entry. 

In U-Ring Paxos, as the last acceptor in the ring is the process that checks whether a decision has been reached, each message originated in the last acceptor includes the ids of decided values. This message is carried along the ring until everyone is informed about the decided values. The coordinator can piggyback new proposals on this message before forwarding it. 
%A buffer entry is freed in u-Ring Paxos after the entry's corresponding Phase~2B is received by the last acceptor and the decision is forwarded along the ring. 

\begin{table}[h]
\center
\caption{Message sizes used in experiments and protocol efficiency (for 10 nodes).}\vspace{4mm}
\label{table:mte}
\begin{tabular}{l|c|c} \hline
Protocol            		& Message size & Efficiency \\ \hline
LCR				& 32 kbytes 	& 91\% \\
\bf{U-Ring Paxos}	& \bf{32 kbytes}	& \bf{90.4}\% \\
\bf{M-Ring Paxos}	& \bf{8 kbytes}	& \bf{90}\% \\
S-Paxos	& 32 kbytes	& 31.2\% \\
%Spread			& 16 kbytes	& 18\% \\
PFSB			& 200 bytes	& 4\% \\ 
Libpaxos			& 4 kbytes		& 3\% \\ \hline
\end{tabular}
%}
%}

\end{table}

\subsection{Ring Paxos versus other protocols}
\label{sec:rpaxosothers}

\begin{figure*}
  \begin{center}
    \begin{tabular}{c@{}c}
      \includegraphics[width=\columnwidth]{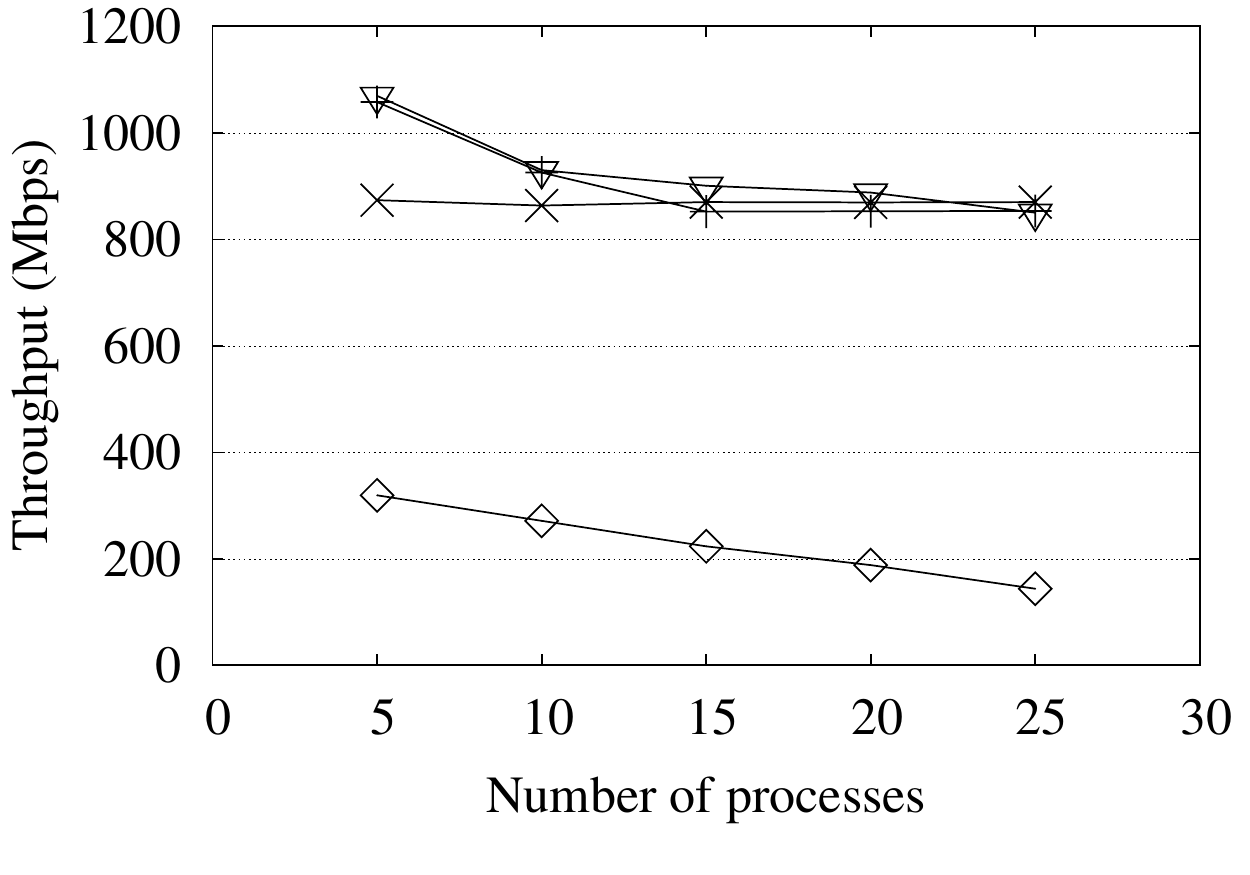} &
      \includegraphics[width=\columnwidth]{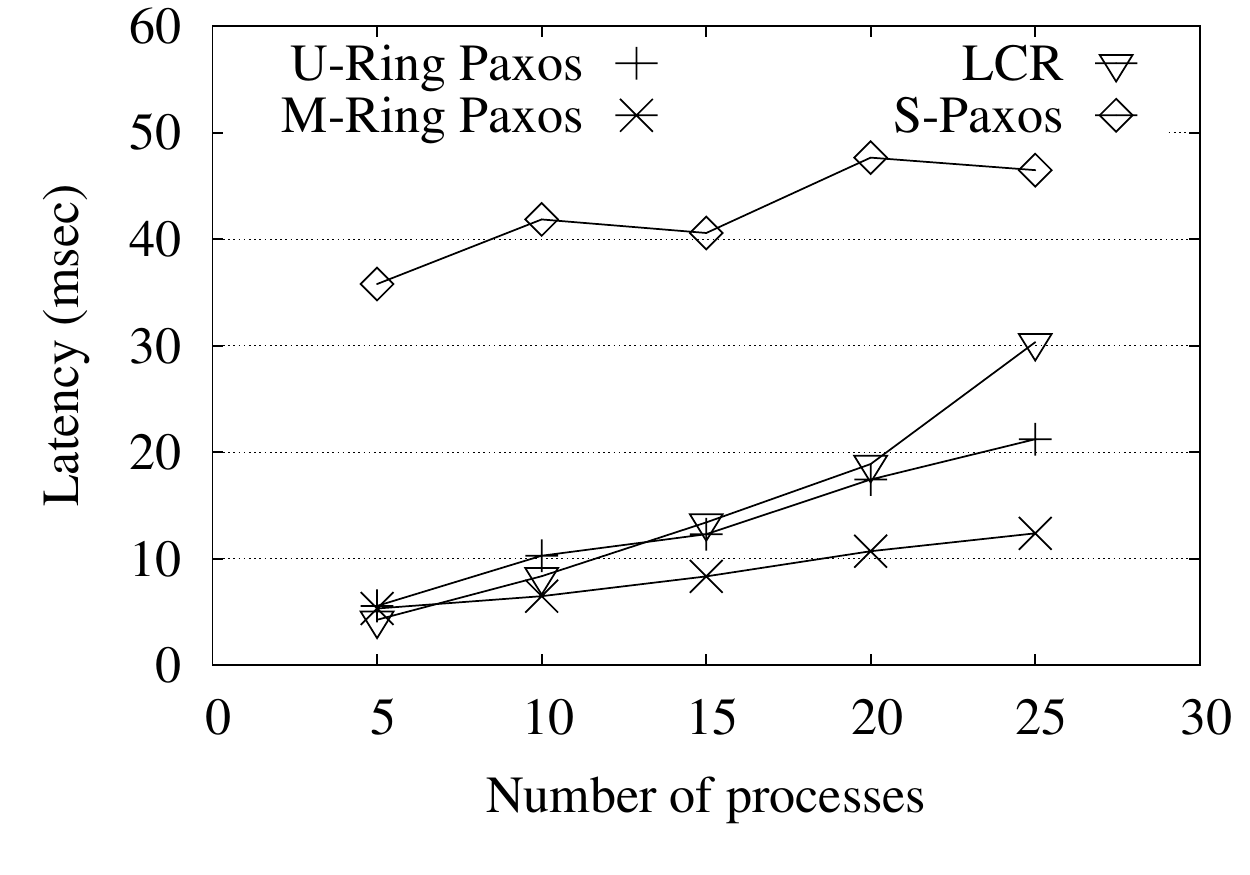} \\

    \end{tabular}
 %   \vspace{-3mm}
    \caption{Throughput and latency when varying the number of processes in the ring.}
    \label{fig:acceptors}
  \end{center}
\end{figure*}

We experimentally compare Ring Paxos to other four atomic broadcast protocols: LCR~\cite{Guerraoui2010}, 
%Spread~\cite{ADM+04}, 
Libpaxos~\cite{Libpaxos}, S-Paxos~\cite{spaxos}, and the protocol presented in \cite{KA08}, which hereafter we refer to as PFSB. 
LCR is a ring-based protocol that achieves very high throughput (see also Section~\ref{sec:rwork}). 
%Spread is one of the most-used group communication toolkits. It is based on Totem~\cite{amir1995totem}. 
Libpaxos, PFSB, and S-Paxos are implementations of Paxos. The first is entirely based on ip-multicast; the second is based on unicast. 
S-Paxos is a unicast-based implementation of Paxos which disseminates the task of receiving and forwarding client requests among all the acceptors.

\begin{figure*}
  \begin{center}
    \begin{tabular}{c@{}c}
      \includegraphics[width=\columnwidth]{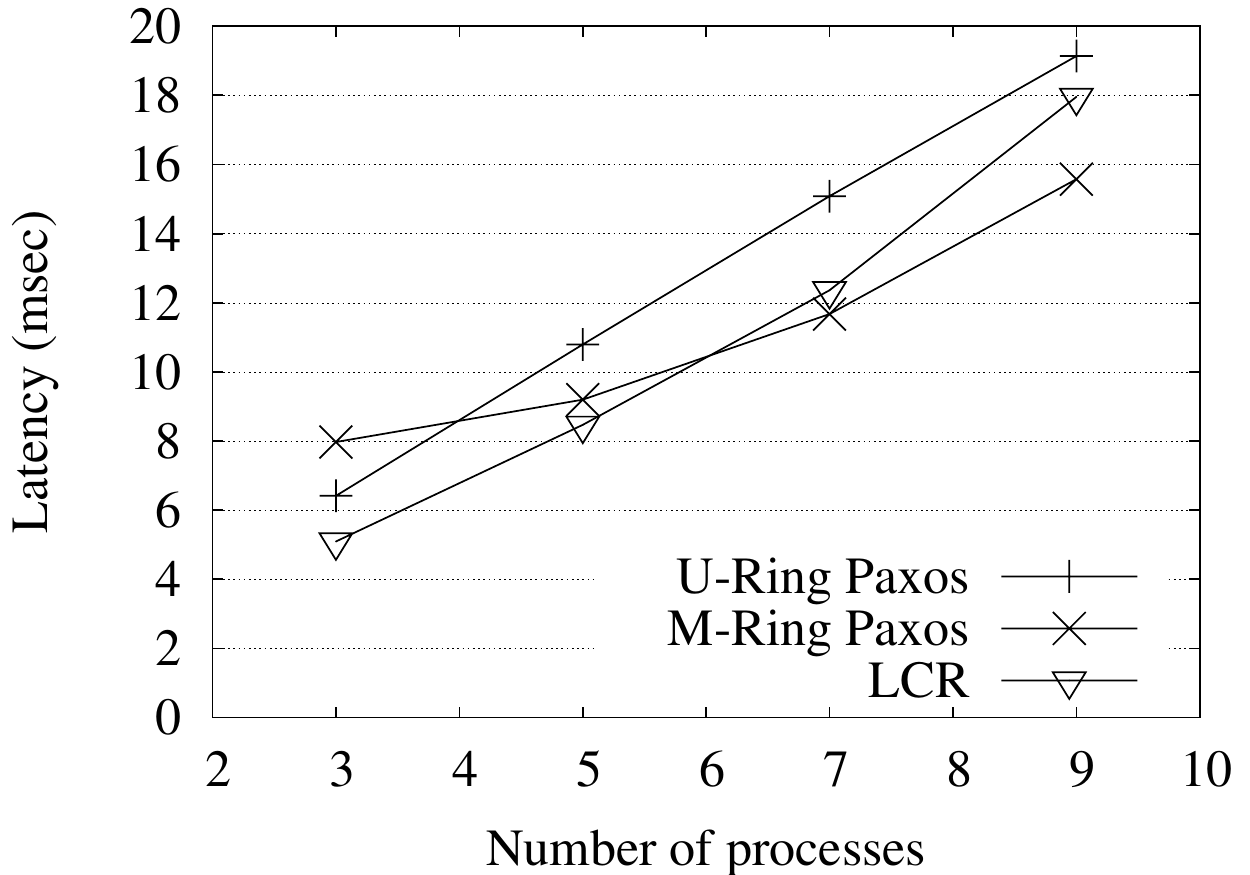} &
      \includegraphics[width=\columnwidth]{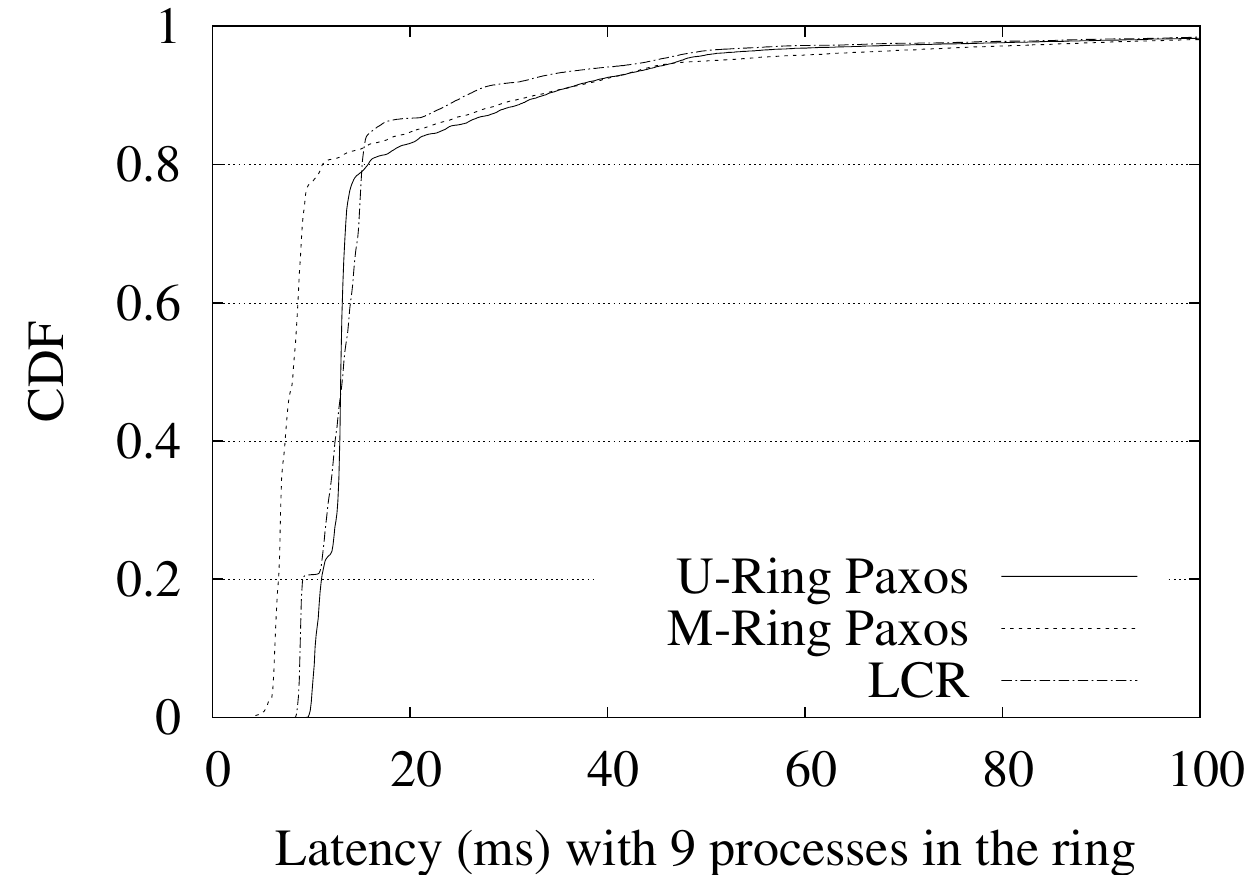} \\
    \end{tabular}
    \vspace{-3mm}
    \caption{Impact of synchronous disk writes on the latency when varying the number of processes in the ring.}
    % We could not evaluate the performance for bigger rings due to the limitations in the number of SSD disks we had. }
    \label{fig:disks}
  \end{center}
\end{figure*}

\begin{figure*}
  \begin{center}

   \begin{tabular}{c@{}c}
      \includegraphics[width=\columnwidth]{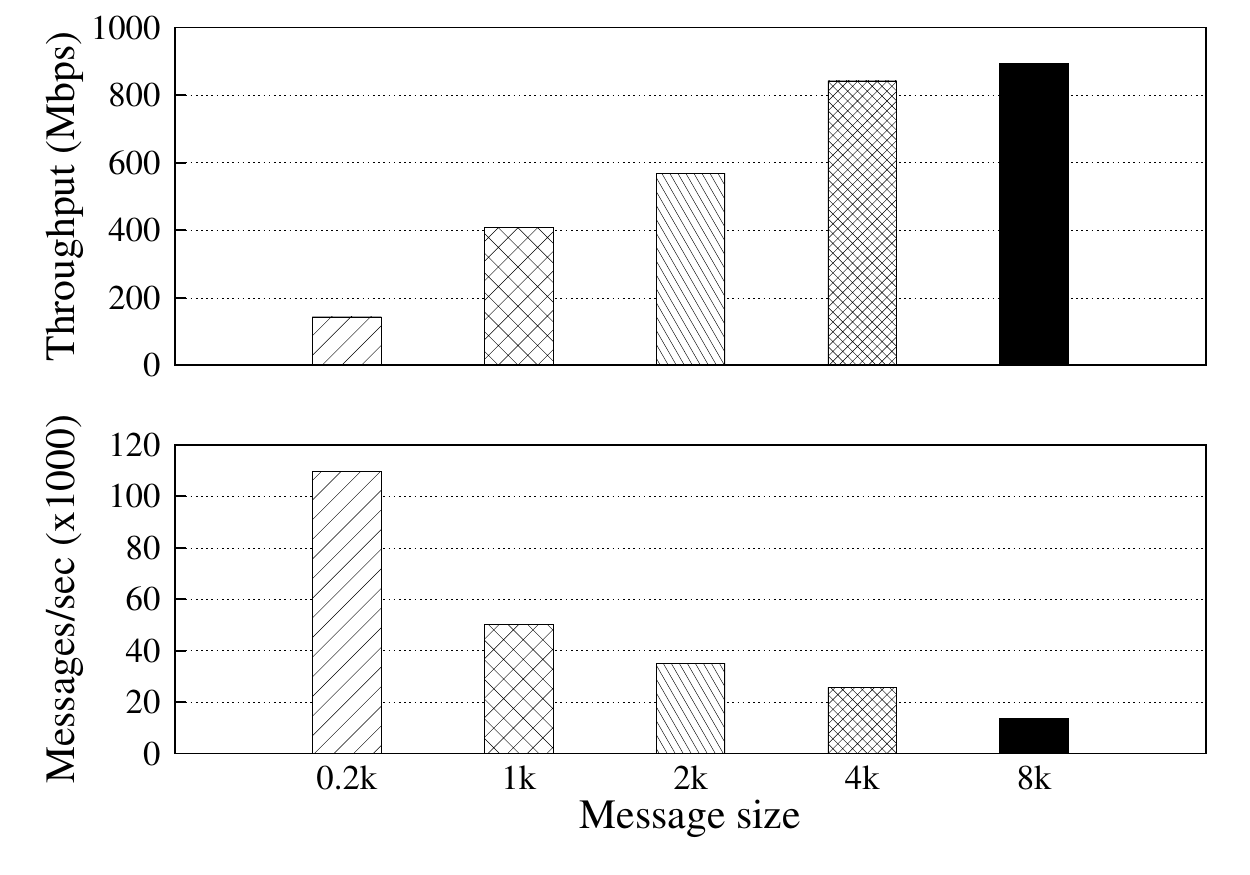} &
      \includegraphics[width=\columnwidth]{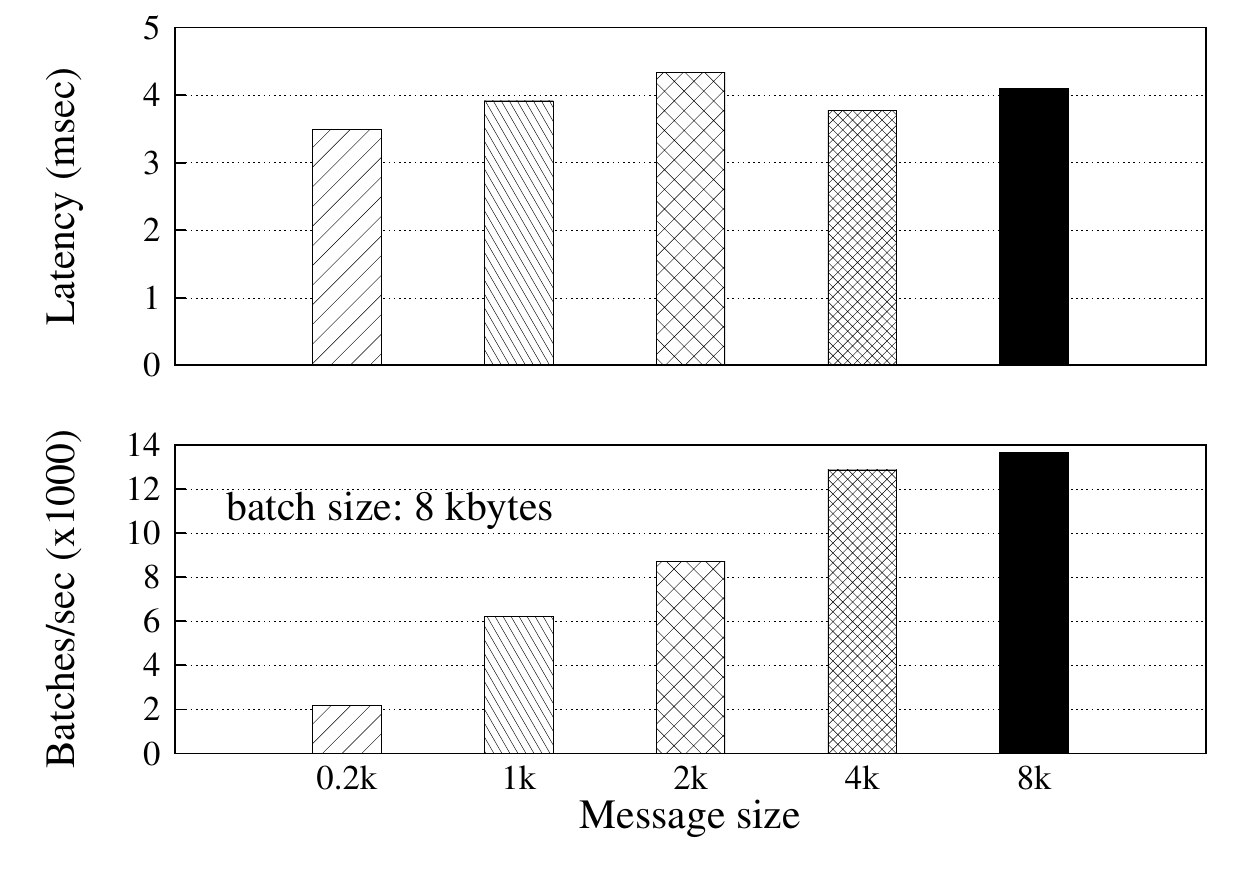}\\
    \end{tabular}
\caption{Impact of application message size on M-Ring Paxos.}
    \label{fig:multicast-msize}

  \end{center}
\end{figure*}

\begin{figure*}
  \begin{center}
    \begin{tabular}{c@{}c}
      \includegraphics[width=\columnwidth]{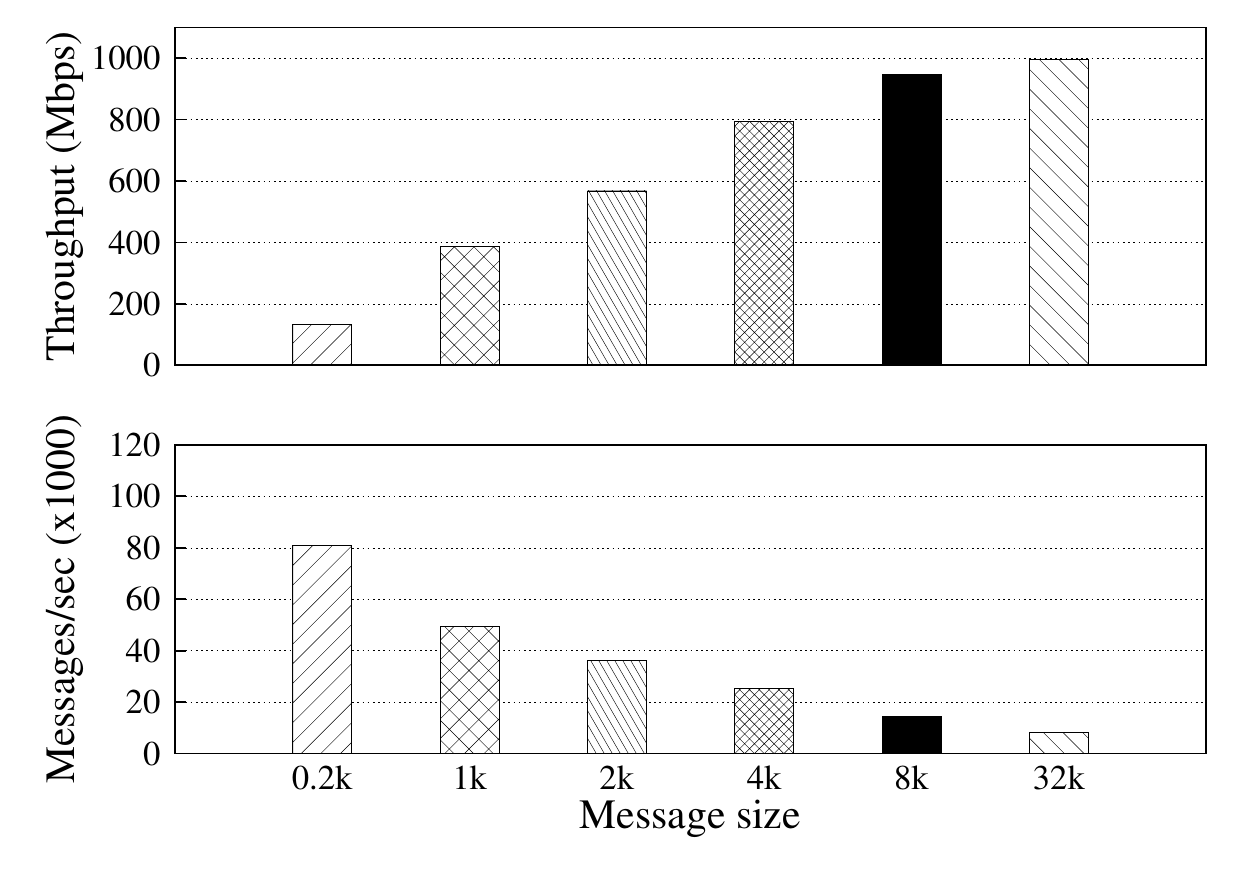} &
      \includegraphics[width=\columnwidth]{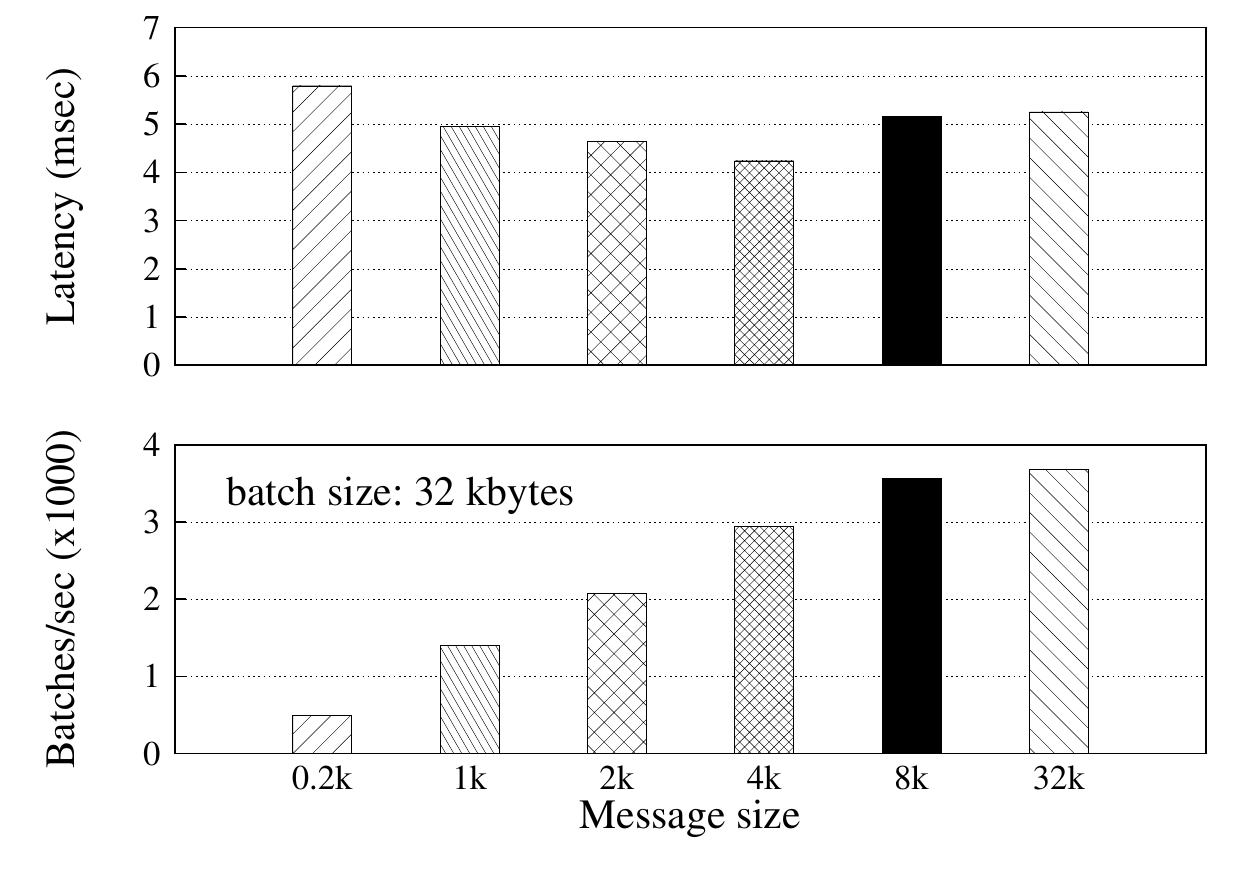}\\
   \end{tabular}
\caption{Impact of application message size on U-Ring Paxos.}
    \label{fig:unicast-msize}

  \end{center}
\end{figure*}

We implemented all protocols, except for PFSB and S-Paxos. 
%We tuned Spread for the best performance we could achieve after varying the number of daemons, number of readers and writers and their locations in the network, and the message size, and some other parameters suggested by the support team of Spread. In the experiments that we report we used a configuration with 3 daemons in the same segment, one writer per daemon, and a number of readers evenly distributed among the daemons.
The performance data of PFSB was taken from \cite{KA08}. The setup reported in \cite{KA08} has slightly more powerful processors than the ones used in our experiments, but both setups use a gigabit switch. The performance data of S-Paxos was obtained with its open-source dissemination. Libpaxos is an open-source Paxos implementation developed by our research group.

Figure~\ref{fig:others} shows the throughput in megabits per second (left graph) and the number of messages delivered per second (right graph) as the number of receivers increases. In both graphs the y-axis is in log scale. 
For all protocols, with the exception of PFSB, we explored the space of message sizes and selected the value corresponding to the best throughput. Table~\ref{table:mte} shows the message sizes used in our experiments. We assess the effect of different message sizes on the performance of Ring Paxos in Section~\ref{subsec:msgsize}.

%The graph on the left of Figure~\ref{fig:others} roughly places protocols into two distinct groups, one group at the top of the graph and the other group at the middle of the graph. The difference in throughput between protocols in the two groups is about one order of magnitude. 

As it is seen in the graph on the left of Figure~\ref{fig:others} protocols based on a ring only (LCR and U-Ring Paxos), on ip-multicast (Libpaxos), and on both (M-Ring Paxos) present throughput approximately constant with the number of receivers. 
%Because it relies on multiple ip-multicast streams, however, Libpaxos has lower throughput. 
%But because of the number of messages exchanged before a request gets ordered their cpu consumption is higher than Ring Paxos protocols. (no evidence! only the number of messages exchanged can be calculated....)

\subsection{Impact of processes in the ring}
\label{sec:procring}

We now consider how the number of processes affects the throughput and latency of the Ring Paxos protocols, LCR, and S-Paxos. 
In Figure~\ref{fig:acceptors}, the x-axis shows the number of acceptors in M-Ring Paxos, U-Ring Paxos, and S-Paxos. In U-Ring Paxos every acceptor is also a proposer and a learner. 
%In the bottom two graphs, the x-axis shows the number of processes. In these experiments there are three acceptors in the rings of M-Ring Paxos and U-Ring Paxos; the rest of the processes are learners and proposers. 
LCR does not distinguish process roles and requires all processes to be in the ring. 

%\begin{figure*}[h]
%  \begin{center}
%    \begin{tabular}{c@{}c}
%      \includegraphics[width=0.5\columnwidth]{graphs/comparison-num-acceptors/graphs/stacked_thr.pdf} &
%      \includegraphics[width=0.5\columnwidth]{graphs/comparison-num-acceptors/graphs/stacked_lat.pdf} \\
%    \end{tabular}
%    \vspace{-3mm}
%    \caption{Varying the number of processes.}
%    \label{fig:acceptors}
%  \end{center}
%\end{figure*}
\begin{figure*}
          \vspace{-15mm}
  \begin{center}
  
    \begin{tabular}{cc}
      \includegraphics[width=\columnwidth]{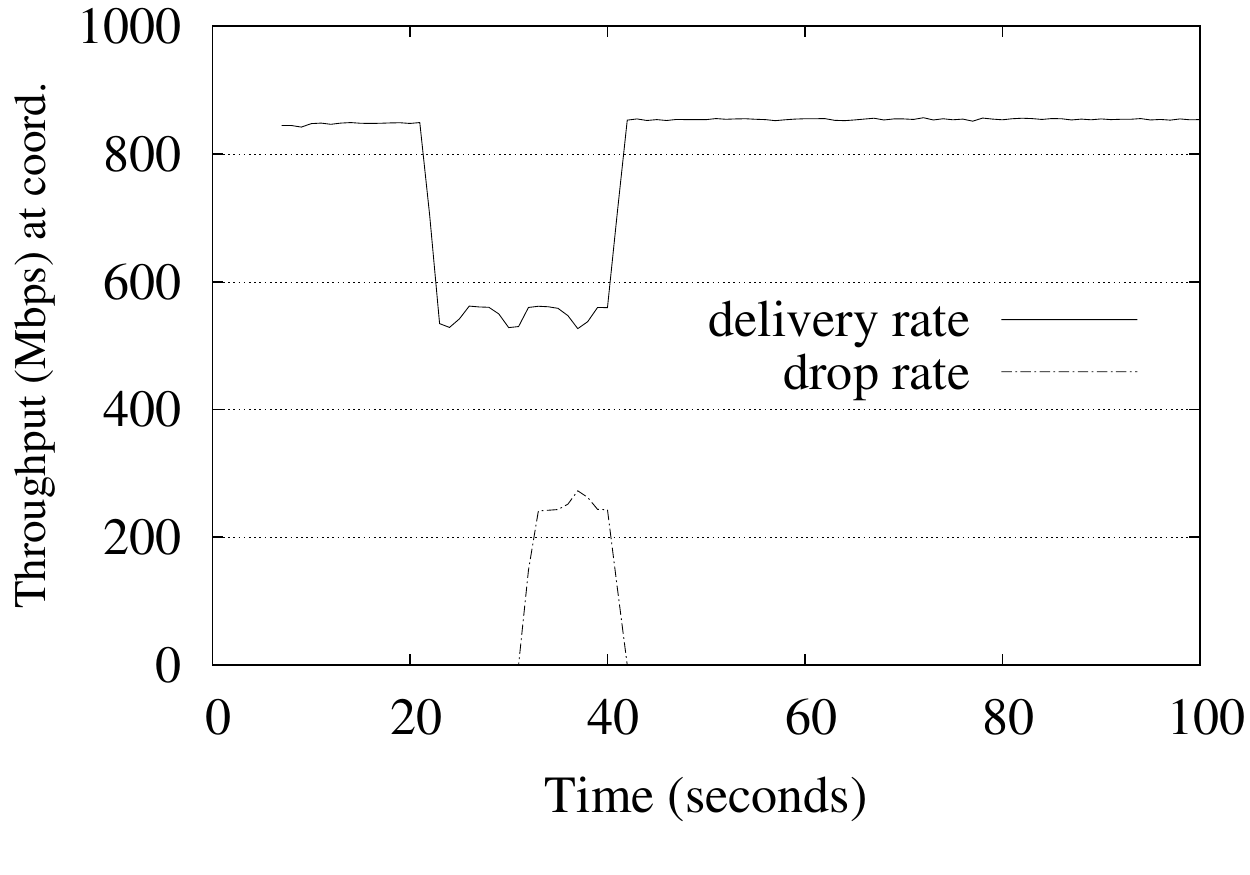} &
       \includegraphics[width=\columnwidth]{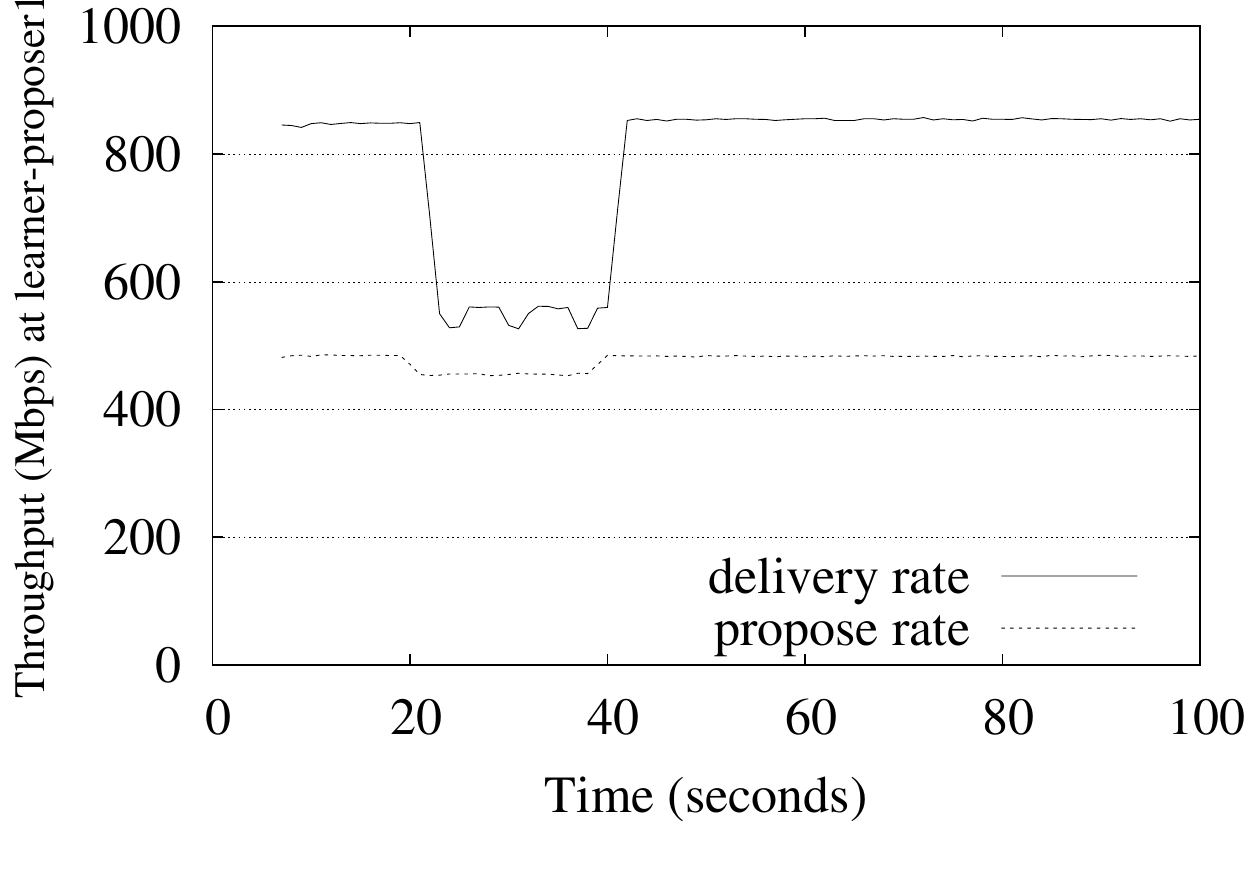} \\
      \includegraphics[width=\columnwidth]{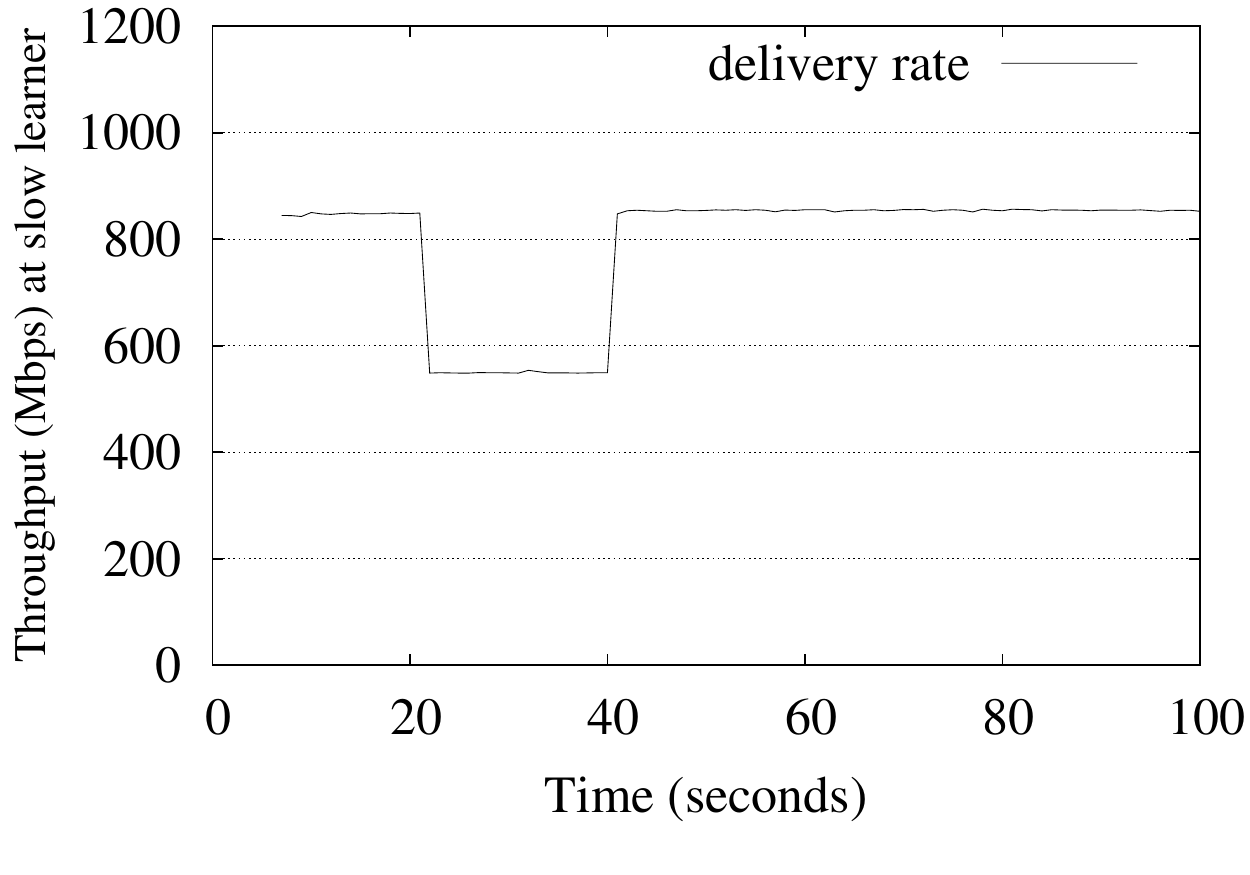} &
      \includegraphics[width=\columnwidth]{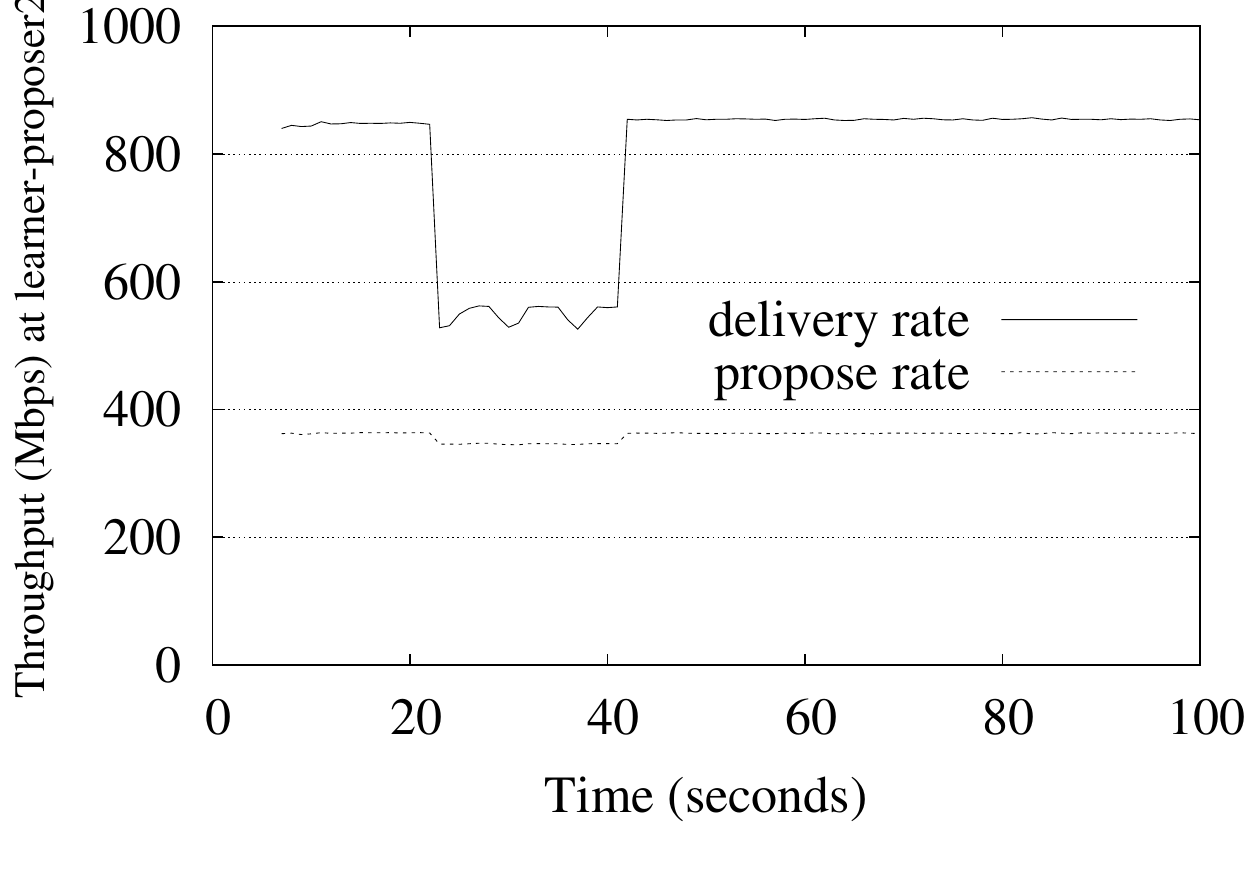} \\

    \end{tabular}
    %\vspace{3mm}
    \caption{Flow control in M-Ring~Paxos. The two learners on the right graphs are also proposers. The learner on the left bottom graph slows down after 20 seconds and returns to its original rate after 40 seconds. }
    \label{fig:fcontrol}
        \vspace{-5mm}
  \end{center}
\end{figure*}

M-Ring Paxos has constant throughput with the number of processes in the ring. Throughput of LCR and U-Ring Paxos slightly decreases as processes are added. With few processes, LCR and U-Ring Paxos can achieve efficiency greater than one, which may look counterintuitive. This happens because in a ring with $n$ processes, $1/n$ of the messages delivered by a process are created by the process itself. Thus, the process can use its available incoming bandwidth to receive messages broadcast by the other processes~\cite{Guerraoui2010}.
In order for LCR and U-Ring Paxos to achieve high throughput, every process in their ring must broadcast messages. 
M-Ring Paxos does not have this constraint.

Figure~\ref{fig:acceptors} shows the latency measured by the message's proposer. 
In U-Ring Paxos and LCR, latencies vary according to the location of the proposer in the ring. 
The values reported for these two protocols are for the best-located proposer, that is, for the proposer with the lowest latency. 

Latency in LCR and U-Ring Paxos degrades with the number of processes; M-Ring Paxos presents a less-pronounced increase in latency as more acceptors are placed in the ring (top right graph in Figure~\ref{fig:acceptors}). Notice that there is less information circulating in the ring of M-Ring Paxos than in the rings in LCR and U-Ring Paxos. In LCR and U-Ring Paxos, the content of each message is sent $n-1$ times, where $n$ is the number of processes in the ring. Message content is propagated only once in M-Ring Paxos (using ip-multicast). 

LCR and U-Ring Paxos present similar latency in Figure~\ref{fig:acceptors}, despite the expected difference, as presented in Table~\ref{table:algo_comparison}. The reason is that for a given setup with $n$ processes, LCR can tolerate $f=n-1$ failures, while U-Ring Paxos can tolerate $f=(n-1)/2$ failures, for an odd $n$. Thus, for the same $n$, the number of communication steps for each protocol is, respectively, $2(n-1)$ and $2.5(n-1)$.

In S-Paxos experiments as we observed substantial variability in the results, due to java's garbage collection mechanism, average values for all experiments were above 35 ms. 
%We do not report the latency for S-Paxos in the graphs as the latency is subject to garbage collection of java. But the minimum average we observed is 35 ms. 

Figure~\ref{fig:disks} shows the performance of M-Ring Paxos, U-Ring Paxos, and LCR when processes store accepted values on disk. All the techniques are essentially disk bound with constant throughput of almost 270 Mbps, regardless the number of processes. However latency increases as nodes are added to the ring. The right-most graph shows the CDF for latency when there are 9 processes in the ring. 
LCR and U-Ring Paxos have comparable latency. M-Ring Paxos has lower latency than LCR and U-Ring Paxos as processes write their values on disk in parallel; in LCR and U-Ring Paxos disk writes across processes happen sequentially. 
%In all the protocols there is only one proposer. 
%The size of the data written to the disk is 32K in all the protocols. 
%In M-Ring Paxos an instance is a batch of 4 values of size 8K. 
In all protocols, data is written on disk in units of 32 Kbytes.

\subsection{Impact of message size}
\label{subsec:msgsize}

Figures~\ref{fig:multicast-msize} and \ref{fig:unicast-msize} quantify the effects of application message size (payload) on the performance of M-Ring Paxos and U-Ring Paxos respectively. In both figures throughput (top left graphs) increases with the size of application messages, up to 8 Kbytes and 32 Kbytes in M-Ring Paxos and U-Ring Paxos, respectively, after which it decreases. Notice that in our prototype ip-multicast packets are 8 Kbytes long, but datagrams are fragmented since the maximum transmission unit (MTU) in our network is 1500 bytes. In U-Ring Paxos communication is based on TCP. Latency is less sensitive to application message size (top right graphs). 

Figures~\ref{fig:multicast-msize} and \ref{fig:unicast-msize} also show the number of application messages delivered as a function of their size (bottom left graphs). Many small application messages can fit a single paxos message and Phase 2 is executed for a batch of proposed values. As a consequence, many application messages can be delivered per time unit (left-most bars). Small messages, however, do not lead to high throughput since they result in high overhead (bottom right graphs).

\subsection{Flow control}
\label{sec:exp:flow-control}

Figure~\ref{fig:fcontrol} illustrates the flow control mechanism with three learners and two proposers using M-Ring Paxos. 
The aggregate proposing rate of proposers is 850 Mbps (right graphs). 
After 20 seconds one of the learners (left bottom graph) slows down. 
As soon as the number of pending instances reaches a predefined threshold, the slow learner sends a notification to one of the acceptors in the ring. 
This acceptor propagates the notification along the ring until it reaches the coordinator. 
Having received this, the coordinator reduces its proposing rate and, as a consequence, the delivery rate drops. 
As proposers keep submitting requests at a constant rate, eventually the receiving buffer of the coordinator overflows and requests are dropped. 
Proposers continue submitting new requests and also re-submitting pending requests (proposers reduce their proposing rate if they detect buffer overflows). 
After 40 seconds the slow learner restores its original rate. 
The coordinator does not receive any slow-down requests and restores its original proposing rate.

%\begin{figure*}[h]
%  \begin{center}
%    \begin{tabular}{cc}
%      \includegraphics[width=0.5\columnwidth]{graphs/flow-control/graphs/fcontrol_1.pdf} &
%       \includegraphics[width=0.5\columnwidth]{graphs/flow-control/graphs/fcontrol_2.pdf} \\
%    \end{tabular}
%    \caption{Flow control in M-Ring~Paxos. The two learners in the right graphs are also proposers. The learner at the left bottom graph slows down after 20 seconds and returns to its original rate after 40 seconds. }
%    \label{fig:fcontrol}
%  \end{center}
%\end{figure*}

\subsection{CPU and memory usage}
\label{sec:cpumem}

Table~\ref{table:mringpaxos} shows the CPU and memory usage of M-Ring Paxos under peak throughput. In the experiments we isolated the processes running M-Ring Paxos in a single node and measured their CPU and memory usage. Not surprisingly, the coordinator is the process with the maximum load since it should both receive a large stream of values from the proposers and ip-multicast these values. 

Table~\ref{table:uringpaxos} shows the results for U-Ring Paxos. In this case, all the processes play the roles of proposer, acceptor and learner. Therefore, the CPU and memory usage of all of them are similar. In both tables memory consumption at coordinator, acceptors and learners is mostly used by the circular buffer of proposed values. For efficiency, in our prototype the buffer is statically allocated.
As a reference, the average CPU usage per process in LCR is in the range of 65\%--70\% and for S-Paxos it is about 270\% (i.e. S-Paxos is multithreaded).
\subsection{Discussion}
\label{sec:exp:con}
In the following, we summarize the main conclusions from the experiments.

\begin{itemize}

\item Judicious use of ip-multicast, as in M-Ring Paxos, and a topology entirely based on a ring, as in U-Ring Paxos and LCR, can achieve throughput near the limits of the network (Section~\ref{sec:rpaxosothers}).

\item Latency increases with the size of the ring in the Ring Paxos protocols, although M-Ring Paxos is less prone to the effects of ring size on latency (Section~\ref{sec:procring}).

\item Both M-Ring Paxos and U-Ring Paxos are susceptible to message size. M-Ring Paxos achieves high throughput with messages of 4 Kbytes or bigger; U-Ring Paxos reaches maximum performance with 8-Kbyte messages.
If application messages are small, batching can improve performance (Section~\ref{subsec:msgsize}).

\item In-memory deployments of M-Ring Paxos and U-Ring Paxos are network-bound; the performance of on-disk deployments is determined by the throughput sustained by the storage device (Section~\ref{subsec:msgsize}).

%\item As shown by our experiments the size of instances play an important role in the performance of Ring Paxos protocols. In case the size of application messages are small, batching as a well known approach, can be used to execute protocol for a set of application messages rather than individual messages and thus preserve the performance of Ring Paxos protocols.

%\item To make a protocol immune to failures, data must be persisted to disk. Although disk usage reduces the performance of our ring-based protocols, the achieved throughput is still better than many other disk-less implementations of Paxos. Despite this, in another work we have addressed the performance issue of disk writes using multiple instances of Ring Paxos protocol~\cite{marandi2012multi}.

\item Our simple flow control mechanism in M-Ring Paxos proved effective in avoiding message losses by slowing down the rate of proposers (Section~\ref{sec:exp:flow-control}).

%Inclusion of a flow control technique is crucial if the underlying communication protocols do not provide sufficient mechanisms to avoid message losses as is the case with UDP protocol used in M-Ring Paxos. We have demonstrated with an experiment how inclusion of a flow control technique can help to detect possible message losses and address it by slowing down the proposers.

\end{itemize}

\begin{table}
\begin{minipage}[t]{0.5\linewidth} % A minipage that covers half the page
\caption{CPU and memory use for M-Ring Paxos.}\vspace{4mm}
\label{table:mringpaxos}

\begin{tabular}{l|c|c} \hline
Role            & \hspace{5mm}CPU\hspace{5mm}     & \hspace{5mm}Memory\hspace{5mm} \\ \hline
Proposer     & 37.2\% & 90 Mbytes \\
Coordinator & 88.0\% & 168 Mbytes \\
Acceptor     & 24.0\% & 168 Mbytes \\
Learner       & 21.3\% & 168 Mbytes \\ \hline
\end{tabular}
\end{minipage}
\hspace{10mm}
\begin{minipage}[t]{0.5\linewidth}
\caption{CPU and memory use for U-Ring Paxos.}\vspace{4mm}
\label{table:uringpaxos}
\begin{tabular}{l|c|c} \hline
Role            & ~~~~CPU~~~~     & Memory \\ \hline
proposer-acceptor-learner     & 48.0\% & 80 Mbytes \\ \hline
\end{tabular}
\end{minipage}

\end{table}

\section{Conclusion}
\label{sec:final}

This paper presents M-Ring Paxos and U-Ring Paxos, two Paxos-like algorithms designed for high throughput. The protocols are optimized for modern interconnects. In order to show that the techniques used are effective, we implemented both protocols and compared them to other atomic broadcast protocols. Our selected protocols include a variety of techniques typically used to implement atomic broadcast. The experiments revealed that both Ring Paxos protocols have the property of providing almost-constant throughput with the number of receivers, an important feature in clustered environments. The experiments pointed out the tradeoffs with pure ring-based protocols. Protocols based on unicast only or ip-multicast only have low latency, but poor throughput. The study suggests that a combination of techniques can lead to the best of both: high throughput and low latency, under weak synchrony assumptions. 

M-Ring Paxos and U-Ring Paxos are available for download from SourceForge~\cite{Libpaxos} and github~\cite{uringpaxos} respectively.

\section{Acknowledgements}
The authors would like to thank Antonio Carzaniga, Leslie Lamport, and Robbert van Renesse for the valuable discussions about Ring Paxos.\

\bibliographystyle{elsarticle-harv} 
\bibliography{main}

\section*{Correctness Proof}
We provide a proof sketch of the correctness of M-Ring Paxos and U-Ring Paxos.
%More precisely, we show that Ring Paxos implements consensus. Atomic broadcast can be built on top of Ring Paxos as done in Paxos.
We focus on properties (ii) and (iii) of consensus. Property (i)
%, if a process decides $v$ then some process proposed $v$, 
holds trivially from the algorithms.

\begin{props}
\vspace{1mm}
%\hspace{-1mm}(ii) No two processes decide different values.
\hspace{-1mm}(ii) No two processes decide different values.
\end{props}
\vspace{1mm}
\noindent Proof sketch: Let $v$ and $v'$ be two decided values, and $v$-$id$ and $v'$-$id$ their unique identifiers. We prove that $v$-$id = v'$-$id$.

%\paragraph{M-Ring Paxos}
\vspace{1mm}
M-Ring Paxos: Let $r$ ($r'$) be the round in which some coordinator $c$ ($c'$) ip-multicast a decision message with $v$-$id$ ($v'$-$id$).
In M-Ring Paxos, $c$ ip-multicasts a decision message with $v$-$id$ after: (a)~$c$ receives $f$+$1$ messages of the form (Phase~1B, $r$,~*,~*); (b)~$c$ selects the value $\mathit{v_{val}} = v$ with the highest round number $\mathit{v_{rnd}}$ among the set $M_{1B}$ of phase 1B messages received, or picking a value $v$ if $\mathit{v_{rnd}} = 0$; (c)~$c$ ip-multicasts  (Phase~2A,~$r$,~$v$,~$v$-$id$); and (d)~$c$ receives (Phase2B, $r$, $v$-$id$) from the second last process in the ring, say $q$.
When $c$ receives this message from $q$, it is equivalent to $c$ receiving $f$+$1$ (Phase~2B,~$r$,~$v$-$id$) messages 
directly because the ring is composed of $f$+$1$ acceptors. Let $M_{2B}$ be the set of $f$+$1$ phase 2B messages. Now consider that coordinator $c$ received the same set of messages $M_{1B}$ and $M_{2B}$ in a system where all processes ran Paxos on value identifiers. In this case, $c$ would send a decide message with $v$-$id$ as well. Since the same reasoning can be applied to coordinator $c'$, and Paxos implements consensus, $v$-$id = v'$-$id$.$\hfill\square$

%\paragraph{U-Ring Paxos}
U-Ring Paxos: Let $r$ ($r'$) be the round in which the last acceptor $a_l$ ($a_l'$) sends a decision message $m_D$ with $v$-$id$ ($v'$-$id$) along the ring.
The proof for U-Ring Paxos is similar to the proof for M-Ring Paxos since U-Ring Paxos only differs in the way $m_D$ is propagated to the learners and the identity of the process who first sends $m_D$. In contrast to M-Ring Paxos where it is the coordinator that sends $m_D$, in U-Ring Paxos, it is the last acceptor in the ring $a_l$ that initiates the propagation of $m_D$ along the ring. Despite this difference, $a_l$ sends $m_D$ when $a_l$ received, including itself, $f+1$ (Phase~2B,~$r$,~$v$,~$v$-$id$) messages, just as in M-Ring Paxos. Since M-Ring Paxos guarantees that $v$-$id$ = $v'$-$id$, the same holds in U-Ring Paxos.$\hfill\square$
\vspace{1mm}
\begin{props}
(iii)  If one (or more) process proposes a value and does not crash then eventually some value is decided by all correct processes.
\end{props}
\vspace{1mm}
\noindent Proof sketch: The proof is almost identical for M-Ring Paxos and U-Ring Paxos. We note the differences when necessary. After GST, processes eventually select a correct coordinator $c$. $c$ considers a ring $c$-ring composed entirely of correct acceptors, for M-Ring Paxos, and a ring $c$-ring composed entirely of correct proposers, acceptors, and learners, for U-Ring Paxos. Coordinator $c$ sends a message of the form (Phase~1A,~*,~$c$-ring) to the acceptors in $c$-ring. Because after GST, all processes are correct and all messages exchanged between correct processes are received, all correct processes eventually decide some value.$\hfill\square$





%\begin{thebibliography}{00}



%\end{thebibliography}

\end{document}